\documentclass[12pt]{article}

\usepackage[T1]{fontenc}
\usepackage[utf8]{inputenc}
\usepackage[english]{babel}

\usepackage{graphicx,psfrag,epsf,color}
\usepackage{amsmath,amssymb,amsfonts}
\usepackage{hyperref}
\usepackage{array}
\usepackage{bm} 
\usepackage[dvipsnames]{xcolor} 
\usepackage{graphicx} 
\usepackage{psfrag,epsf,color,xcolor}
\usepackage{multirow}
\usepackage{rotating}
\usepackage{slashed}
\usepackage{textcomp}
\usepackage[font=small]{caption} 
\usepackage{float}
\usepackage{comment} 
\usepackage{cite} 
\usepackage{bbold}
\usepackage[normalem]{ulem}  
\usepackage{mathtools}
\usepackage{multicol}
\usepackage[normalem]{ulem}

\newcommand{\secref}[1]{\hyperref[sec::#1]{SECTION~\ref*{sec::#1}}}
\newcommand{\subsecref}[1]{\hyperref[subsec::#1]{SECTION.~\ref*{subsec::#1}}}
\newcommand{\figref}[1]{\hyperref[fig::#1]{FIG.$\,$\ref*{fig::#1}}}
\newcommand{\tabref}[1]{\hyperref[tab::#1]{TABLE$\,$\ref*{tab::#1}}}
\newcommand{\eqnref}[1]{\hyperref[eqn::#1]{Eq.$\,$(\ref*{eqn::#1})}}
\newcolumntype{C}{>{$}c<{$}} 

\newcommand{\E}{\mathrm{e}}
\newcommand{\I}{\mathrm{i}}
\newcommand{\D}{\text{d}}
\newcommand{\arccot}{\text{arccot}}

\newcommand{\fig}{Fig.}

\newcommand{\cf}{\emph{cf.}}

\newcommand{\UV}{ \text{PUV} }
\newcommand{\B}{{\mathcal{B}}}
\newcommand{\eS}{{\mathcal{S}}}
\newcommand{\BSF}{\text{BSF}}

\newcommand{\as}{{\alpha_s}}
\newcommand{\ab}{{\alpha_b}}

\newcommand{\zbb}{\zeta_n}
\newcommand{\zbbpeak}{\zbb^\text{peak}}
\newcommand{\zs}{\zeta_s}
\newcommand{\gbb}{\gamma_n}

\newcommand{\el}{\ell}
\newcommand{\lp}{{\el^{\prime}}}
\newcommand{\mP}{m^{\prime}}

\renewcommand{\vec}[1]{\boldsymbol{#1}}

\newcommand{\lr}{\left(}
\newcommand{\rr}{\right)}
\newcommand{\be}{\begin{equation}}
\newcommand{\ee}{\end{equation}}
\newcommand{\bea}{\begin{eqnarray}}
\newcommand{\eea}{\end{eqnarray}}
\newcommand{\nn}{\nonumber}

\numberwithin{equation}{section}

\begin{document}
\allowdisplaybreaks

\begin{titlepage}

\begin{flushright}
{\small
TUM-HEP-1534/24\\
\today 
}
\end{flushright}

\vskip1cm
\begin{center}
{\Large \bf Perturbative unitarity violation in radiative 
capture\\[1.ex] transitions to dark matter bound states}\\[0.6cm]  
\end{center}

\vspace{0.5cm}
\begin{center}
{\sc Martin Beneke, Tobias Binder, Lorenzo de Ros, Mathias Garny,\\ Stefan Lederer
}\\[0.7cm]
{\it Physik Department T31,\\
James-Franck-Stra\ss{}e~1, 
Technische Universit\"at M\"unchen,\\
D--85748 Garching, Germany}

\end{center}

\vspace{0.6cm}
\begin{abstract}
\vskip0.2cm\noindent
We investigate the formation of bound states of non-relativistic dark matter particles subject to long-range interactions through radiative capture. The initial scattering
and final bound states are described by Coulomb potentials with different strengths, as relevant for non-abelian
gauge interactions or theories featuring charged scalars.
For bound states with generic quantum numbers $n$ and $\ell$, we provide closed-form expressions for the bound-state formation (BSF) cross sections 
of monopole, dipole and quadrupole transitions, and of arbitrary multipole order when $\ell=n-1$. This allows us to
investigate in detail a strong enhancement of BSF that occurs for initial states in a repulsive potential. For $\ell=n-1\gg 1$, we show that the BSF cross section for
each single bound state violates the perturbative unitarity bound in the vicinity of a certain critical initial velocity,
and provide an interpretation in terms of a smooth matching of classical trajectories. When summing the BSF cross section over
all possible bound states in the final state, this leads to a unitarity violation below a certain velocity, but within the validity range
of the weakly coupled non-relativistic description. We identify 
an effectively strong interaction as the origin of this unitarity violation, which is caused by an ``anomalously'' large overlap of scattering and bound-state wave functions in Coulomb potentials of different strength.

\end{abstract}
\end{titlepage}

\thispagestyle{empty}
\tableofcontents
\newpage
\setcounter{page}{1}

\section{Introduction}
\label{sec:intro}

The time-honoured subject of non-relativistic quantum-mechanical 
scattering and bound states has recently seen an unexpected application 
to the physics of dark-matter annihilation. Dark matter (DM) particles 
that can exchange a lighter particle experience a long-range force, 
which significantly modifies the scattering and pair annihilation 
cross sections and forms dark-matter bound states, if the force 
is sufficiently attractive (``Sommerfeld-effect''). 
For Yukawa forces generated by the 
exchange of a massive particle, the cross sections can be resonantly 
enhanced by orders of magnitude for particular mass values 
\cite{Hisano:2003ec,Hisano:2004ds}, reflecting the 
universal physics of large 
scattering length \cite{Braaten:2013tza} in the presence of a shallow 
bound state (or quasi-bound state for higher partial waves 
\cite{Beneke:2024iev}). The spectrum of bound states itself can play an 
important role for the annihilation dynamics 
\cite{vonHarling:2014kha,An:2016gad,Asadi:2016ybp} due to radiative capture before 
annihilation, and transitions among the bound-state levels.

These phenomena are far from exotic. They occur for the standard 
electroweakly charged WIMP when its mass is of order TeV or larger, and in dark-sector models that contain 
a lighter mediator particle, and therefore potentially in a variety 
of interactions that goes beyond those of quantum electrodynamics (QED), 
nuclear physics and condensed matter/atomic systems, for which such effects 
have previously been studied. This raises interesting fundamental 
questions. Since the mass of the DM particle is not known, the scattering length of the system may be very large near certain values of the mass, 
in which case the scattering and annihilation cross sections 
violate the perturbative quantum-mechanical unitarity bound. In 
this case, unitarity is restored by including the annihilation 
cross section as a contact interaction into the Schr\"odinger 
equation~\cite{Blum:2016nrz,Parikh:2024mwa} (see~\cite{Flores:2024sfy} for a generalization to non-contact $2\to 2$ interactions). This can be understood in analogy with 
nucleon-nucleon scattering \cite{Kaplan:1998we} by the fact 
that at very low velocity of the DM particles, the Yukawa potential 
becomes a contact interaction, which flows to a fixed-point 
at which the contact interaction becomes relevant in the 
non-relativistic effective theory, in contrast to the 
standard non-relativistic counting.

Our concern here is the recent observation 
\cite{Oncala:2019yvj, Binder:2023ckj} 
that perturbative unitarity violation ($\UV$) also occurs in 
the radiative capture process
\begin{equation}
\label{eq:process}
\mathcal{S}\rightarrow\mathcal{B}+\gamma
\end{equation}
from a DM two-particle scattering state into a bound state by emission of a massless particle, denoted generically as $\gamma$. Different 
from QED, where unitarity is respected, the relevant situation 
is when the attractive potential of the bound state, $V_b$, 
differs from the scattering-state potential $V_s$.

As any inelastic $2\to 2$ process, the cross section for radiative 
capture from a scattering state of distinguishable particles 
must respect the partial-wave unitarity bound~\cite{Griest:1989wd} 
\begin{equation}
\label{eq:unitaritybound}
(\sigma v)_{\mathcal{S}_\lp\rightarrow\mathcal{B}+\gamma} 
\leq (\sigma v)^\text{uni}_{\lp} 
\equiv \frac{4\pi}{M^2}\frac{2\lp+1}{ v}\,,
\end{equation}
where $M$ denotes the mass of the DM particle, and $ v$ the 
relative velocity of the two particles in the scattering state 
$\mathcal{S}$.
However, recently \UV was found for the following two cases:
\begin{itemize}
\item Capture of two scalar particles into a bound state via monopole emission of another charged scalar \cite{Oncala:2019yvj}. 
\item Capture of coloured particles via dipole emission of a gluon 
\cite{Binder:2023ckj}. Unitarity was found to be systematically 
violated even at arbitrarily small couplings, if the capture rate 
was summed over sufficiently many excited bound states $n$. 
\end{itemize}
In both  cases, the scattering state experiences a different, repulsive potential from the bound state. We note that \UV is expected generally in effective 
field theories when the scattering momenta approach and exceed 
the limit of validity of the effective description, and more 
generally at strong coupling, when the perturbative expansion 
breaks down. It is therefore important to emphasize that the 
above two observations of \UV in radiative capture occur long 
before the onset of strong coupling. The origin of this \UV must 
therefore be sought in the {\em dynamical} generation of a 
new scale (as in the case of large scattering length discussed 
above) or an {\em effectively} large coupling.

The above observations of \UV at weak coupling are purely 
numerical. In the present work, we take a closer look at 
the analytic expressions for the overlap matrix elements and 
capture cross sections for arbitrary (``electric'') multipole 
transitions and identify the parameter ranges of \UV 
analytically in certain limits. Unitarity is already violated for the exclusive 
transition to the single final state $n=\el+1\gg1$, in which 
case the analytic results are particularly simple.  
For this case we demonstrate an anomalous enhancement of the 
wave-function overlap as the cause of unitarity violation 
and provide a semi-classical interpretation: the capture happens 
precisely when the classical scattering trajectory grazes the 
bound orbit, and the particle transits into the bound orbit 
by releasing the energy difference into the radiated 
boson.

This work is organized as follows: In Sec.~\ref{sec:Setup} we review the formalism for radiative bound-state formation in the non-relativistic limit, and we provide closed-form results for the cross section suitable for efficient numerical evaluation for arbitrary bound-state levels and applicable to Coulomb potentials of different strength for the initial and final state in Sec.~\ref{sec:Overlap}. We analytically demonstrate the enhancement of the cross section observed for repulsive initial potentials in Sec.~\ref{sec:UV}, for bound states with maximal angular momentum and transitions of any multipole order. We then develop a semi-classical and intuitive interpretation of the strong enhancement in terms of an ``anomalously'' large wave function overlap in Sec.~\ref{sec:class}, and generalize the discussion to bound states of arbitrary angular momentum. We discuss how the ``anomalous'' enhancement of the radiative bound-state formation cross section can lead to a violation of perturbative unitarity bounds in Sec.~\ref{sec:puv}, and conclude in Sec.~\ref{sec:conclusions}. Appendix~\ref{app:Radial} contains details on the computation of the radial overlap integral as well as further results applicable to general multipole transitions, and Appendix~\ref{sec:Airy} a derivation of the semi-classical approximation to the bound-state formation cross section.

\section{Radiative capture} 
\label{sec:Setup}

We are interested in the capture of two dark-matter particles 
of mass $M$ from a scattering eigenstate $|\vec{p}\rangle$ into a 
bound state $|n\el m\rangle$ via emission of a massless particle 
$\gamma$ with momentum $\vec{k}$ and energy $\omega = |\vec{k}|$. 

The bound states are characterized by the usual principal quantum 
number $n$, the angular momentum $\ell$ and its projection 
along the $z$-axis, $m$. Furthermore, we assume them to be 
non-relativistic and bound by an attractive Coulombic potential, 
that is, they are eigenstates of the Schr\"odinger operator
\begin{equation}
\label{eq:HbVb}
H_{b} = -\frac{\vec{\nabla}^2}{2\mu}+V_{b}(r), \qquad 
V_{b}(r) = -\frac{\alpha_{b}}{r}\,,
\end{equation}
with generic effective strength $\alpha_b>0$ and reduced mass $\mu=M/2$. The scattering states with momentum $\vec{p}$ and angular momentum $\lp \mP$ are 
also described by a Coulomb potential,
\begin{equation}
\label{eq:Vs}
  V_s(r) = -\frac{\alpha_s}{r}\,.
\end{equation}

For this work, it is important that the effective coupling strength 
$\alpha_s$ differs from $\alpha_b$, including the possibility of a 
repulsive potential if $\alpha_s$ is negative. We denote their ratio 
by
\begin{equation}
  \kappa \equiv \frac{\alpha_s}{\alpha_b}\,.
\end{equation}
As an example, consider an unbroken SU(N) gauge interaction with gauge coupling $g$ and interaction strength $\alpha=g^2/(4\pi)$. If DM particles transform under its fundamental representation, DM particle anti-particle states can be 
classified by their total ``colour'' charge, either in the singlet (neutral) 
or the adjoint representation. Bound states can exist only in the singlet configuration with $\alpha_b=C_F\,\alpha=(N^2-1)/(2N)\alpha>0$, while the scattering states can  form either in the singlet or the adjoint. For bound-state formation via emission of an SU(N) gauge boson $\gamma$, colour conservation demands the scattering state to be in the adjoint, with repulsive potential characterized by $\alpha_s=(C_F-C_A/2)\alpha=-1/(2N)\alpha<0$. In this case, 
\be
  \kappa=1-\frac{C_A}{2C_F}=-\frac{1}{N^2-1}<0\,.
\label{eq:kappa}
\ee
For a U(1) interaction, one instead recovers the QED case $\kappa=1$, since the gauge boson $\gamma$ is neutral. These scenarios apply to dark sector models featuring a weakly coupled dark non-abelian or abelian gauge interaction, respectively\cite{Harz:2018csl,Binder:2020efn,Binder:2021otw,Biondini:2023zcz}. Additionally, they also apply to models with coloured or charged mediator particles. In that case, the bound and scattering states are formed from particle-antiparticle pairs of mediators, and the gauge interaction can be identified with the usual Standard Model QCD or $U(1)_Y$, respectively~\cite{Biondini:2018ovz,Gross:2018zha,Garny:2021qsr,Becker:2022iso,Binder:2023ckj,Becker:2023opo}.

The leading amplitude for the emission of a massless gauge boson 
in the capture process is given by the transition matrix element 
\begin{equation}
\langle n\ell m|\vec{r}|\vec{p}\rangle
\label{eq:E1}
\end{equation} 
of the electric dipole interaction. To establish contact with the underlying 
dark matter field theory model, we first note that the leading coupling of 
a massless vector field $A^\mu$ to non-relativistic particles is always 
described by the non-relativistic effective Lagrangian
\begin{eqnarray}
\label{eq:nreft}
{\cal L}_{\psi}&=&
\sum_{i=\rm DM,\overline{DM}} 
\psi_i^{\dagger} \bigg( \I D^0 + \frac{{\bf D}^2}{2M}\bigg)\psi_i\,,
\end{eqnarray}
where $\psi_i$ denote non-relativistic, scalar 
dark-matter fields.\footnote{Since the coupling to the emitted particle is 
spin-independent, the conclusions of this paper are independent of the 
spin of the dark-matter particle.} In a physical gauge, where 
$A^0$ is non-dynamical, the single emission amplitude originates from 
the gauge-invariant kinetic term and reads 
\begin{equation}
\frac{g}{2M}\left( (\I \vec{\nabla}\psi)^\dagger T^a \psi+\psi^\dagger T^a \I \vec{\nabla}
\psi\right)\vec{A}  \equiv -\vec{j}^a\vec{A}^a\,, 
\end{equation}
where $a$ denotes an adjoint SU(N) index (absent for U(1)) and $T^a$ the 
generator in the fundamental representation.

It is well-known that the emitted radiation field 
can be classified according to its angular momentum when the wavelength 
is larger than the size of the source, as is always the case for 
radiative capture, and expanded into a series of electric and 
magnetic multipole fields. In order to be able to investigate PUV 
for various transitions, we generalize \eqref{eq:E1} to electric  
multipole transitions of arbitrary order $L$, for which the 
matrix element reads 
\begin{equation}
Q_{LM,fi}\equiv 
\langle n\ell m|r^LY_{LM}^*(\hat{\vec{r}})|\vec{p}\rangle 
= \int d^3\vec{r} \,{\cal B}_{n\ell m}^*(\vec{r}) 
r^LY_{LM}^*(\hat{\vec{r}}){\cal S}_{\vec{p}}(\vec{r})\,,
\label{eq:EL}
\end{equation} 
where ${\cal S}_{\vec{p}}(\vec{r})$  and  $ {\cal B}_{n\ell m}(\vec{r})$
denote the scattering initial-state ($i$) and bound final-state ($f$) 
Coulomb wave functions in 
coordinate space, respectively,  given in App.~\ref{app:wave functions}.  
$Y_{LM}(\hat{\vec{r}})$ refers to the standard spherical harmonics 
and $\hat{\vec{r}} = \vec{r}/r$. 
The capture cross section for fully specified quantum numbers of 
the bound and scattering states as well as a fixed multipole order $L$, but summed over the two polarization states 
of the emitted particle and over $M$, and integrated over the three-momentum $\vec{k}$ of the emitted particle, 
is given by~\cite{Blatt:1952ije}
\begin{equation}
(\sigma v)_{i\to f}^{(L)}  = g^2\frac{2 (L+1)}{L [(2L+1)!!]^2}\,
 \omega^{2L+1}\, \sum_{M=-L}^L \left|Q_{LM,fi}\right|^2\,,
\label{eq:ELexclusivecrossection}
\end{equation}
where the photon energy $\omega$ is determined by 
$\omega=\vec{p}^2/(2\mu)-E_n$ (and $E_n\equiv -\alpha_b^2\mu/(2n^2)$).
Since the overlap integral \eqref{eq:EL} is governed by $r$ of order of 
the Bohr radius of the bound state, increasing the multipole order 
by one suppresses the capture cross section by two powers of $r\omega$, corresponding parametrically to a factor $\alpha_b^2$.

In practice, one is interested in transitions to bound states summed over 
$m=-\ell,\ldots, \ell-1,\ell$ and from scattering states  
integrated over the direction of its relative momentum, 
$\hat{\vec{p}}$. The integration can be done by writing 
\begin{equation}
\eS_{\vec{p}}(\vec{r}) = 
\sum_{\lp=0}^\infty \eS_{p\lp}(r)\, 4\pi\, \I^{\lp} \E^{\I \,\text{arg}\lr\Gamma(1+\lp-\I\zs)\rr} \sum_{\mP=-\lp}^{\lp} Y^*_{\lp \mP}(\hat{\vec{r}})Y_{\lp \mP}(\hat{\vec{p}})\,,
\label{eq:spfact}
\end{equation}
which factorizes the dependence 
on $\vec{p}\cdot\vec{r}$ and defines the partial-wave components of the scattering 
wave function.  It is convenient to use the notation
\be
\label{def:Zeta}
\zbb \equiv \frac{\ab}{nv},\qquad \zs\equiv 
\frac{\as}{v}\,.
\ee
Note that for any given $n$, only two out of the three variables 
$\zbb,\zs$ and $\kappa$ defined in \eqref{eq:kappa} are independent, 
as $\zs=\kappa n\zbb$, and the photon energy can be expressed as 
\begin{equation}
\omega=(1+\zbb^2)\,\frac{\mu v^2}{2}\,.
\end{equation}
The variable $\zbb$ can also be written as the ratio of the Bohr momentum $p_n=\ab\mu/n$ of the bound state and the relative momentum $p=\mu v$ of the scattering state, $\zbb=p_n/p$.

Introducing \eqref{eq:spfact} in \eqref{eq:EL} allows the integral over 
the direction of $\vec{r}$ to be performed. For given $\ell$ and multipole order $L$, this imposes the familiar angular 
momentum selection rules on the transition. The capture cross section from 
an initial scattering state with relative 
three-momentum $p=\mu v$ (integrated over its direction) into a 
bound state $n\ell$ (summed over $m$) is then given by 
\begin{equation}
(\sigma v)_{p\to n\ell}^{(L)} = \sum_{\lp} 
(\sigma v)_{p\lp\to n\el}^{(L)}
\end{equation}
where the partial-wave cross sections can be expressed 
as
\begin{equation}
 (\sigma v)_{p \lp \to n \el}^{(L)} = g_{L,\rm eff}^2
\frac{4}{9}\,
 \omega^{2L+1}\times I_A\times I_R\,,
\label{eq:svIRIA}
\end{equation}
with 
\begin{equation}
g_{L,\rm eff}^2 = g^2\,\frac{9 (L+1)}{2 L [(2L+1)!!]^2} 
\label{eq:geff_QED}
\end{equation}
defined such that $g_{L,\rm eff}^2=g^2$ for the 
case of electromagnetic dipole transitions, $L=1$.

The integral over the direction of $\vec{r}$ can be reduced to a well-known integral over three Legendre polynomials~\cite{Landau:1991wop} and expressed in terms of the Wigner three-$j$ symbols as~\cite{Varshalovich:1988ifq} 
\begin{equation}
  I_A \,=\, (2L+1)(2\el+1)(2\lp+1) 
  \begin{pmatrix}
    L & \el & \lp\\
    0 & 0 & 0
  \end{pmatrix}^2 .
\end{equation}
The Wigner $3j$ symbol is non-zero only if the 
triangle inequalities $|\ell-\ell'|\leq L$
and $\ell+\ell'\geq L$ are satisfied. Furthermore, $L+\ell+\ell'$ needs to be even. For the dipole case ($L=1$) this implies the usual selection rule $\ell'=\ell\pm 1$. For monopole transitions ($L=0$, see~\eqref{eq:scalar} below) one has $\ell'=\ell$, and for quadrupole transitions ($L=2$), $\ell'=\ell\pm 2$ or $\ell'=\ell\geq 1$.

One is left with the square of the radial overlap 
integral 
\begin{equation}
\label{def:IR}
  I_R~\equiv~  \left|\int_0^\infty \! dr\,r^2\, {\cal B}_{n\ell}^*(r) \,r^L\,{\cal S}_{p\ell'}(r) \right|^2\,,
\end{equation}
expressed in terms of the bound and scattering state radial wave functions, ${\cal B}_{n\ell}(r)$ and ${\cal S}_{p\ell'}(r)$, respectively. 
Using their explicit forms in terms of ${}_1F_1$ hypergeometric functions for Coulombic potentials, given in  \eqref{eq:radialB} and \eqref{eq:radialS}, we find an explicit result for 
general $L$, $n$, $\ell$ and $\ell'$ in terms of a sum over ${}_2F_1$ hypergeometric functions. The derivation is provided in Appendix~\ref{app:Radial}. The radial integral contains all dynamical information on the capture process and the origin of PUV must be sought in properties of the scattering- and bound-state wave-function overlap for Coulomb potentials of different strength.

The above derivations holds for a general U(1) gauge interaction, which is not the case of interest. The generalization to SU(N) is, however, straightforward. While the bound state 
must be in the ``colour''-singlet state, the scattering 
states must be generalized to $|\vec{p},ij\rangle$, where 
$ij$ denote the SU(N) indices of the DM particle and 
anti-particle in the fundamental representation. The 
corresponding wave function is a 
superposition
\begin{equation}
\mathcal{S}_{\vec{p},ij}(\vec{r}) = 
a\cdot \frac{\delta_{ij}}{\sqrt{N}}\,
S_{\vec{p},\rm sing}(\vec{r})+
b\cdot \sqrt{2}\sum_{A=1}^{N^2-1}T^A_{ij}\,
S^A_{\vec{p},\rm adj}(\vec{r})
\end{equation}
with $\sqrt{a^2+b^2}=1$, where $S_{\vec{p},\rm sing}(\vec{r})$, $S_{\vec{p},\rm adj}(\vec{r})$ are the normalized eigenfunctions that diagonalize the Hamiltonian in ``colour'' 
space. One usually assumes that the DM particles in the 
primordial plasma form an incoherent mixture of random 
colour states of the DM particle and anti-particle 
separately, so that the initial scattering state is described by the density matrix
\begin{equation}
\varrho_{ij,i'j'}=\frac{\delta_{ii'} \delta_{jj'}}{N^2}.
\end{equation}
One then obtains \eqref{def:IR} with ${\cal S}_{p\ell'}(r)
\to {\cal S}_{p\ell',\rm adj}(r)$ and the effective 
coupling for the emission of the ``gluon'' acquires 
a group-theory factor according to 
\begin{equation}
g_{L,\rm eff}^2 \to \frac{C_F}{N^2} \cdot g_{L,\rm eff}^2\,.
\label{eq:geff_QCD}
\end{equation}

The expression \eqref{eq:svIRIA}, together with the effective couplings \eqref{eq:geff_QED} and \eqref{eq:geff_QCD}, applies when the emitted $\gamma$ carries spin 1, hence the multipole order must satisfy $L>0$. When the emitted particle is a scalar, by adapting the  derivation in \cite{Blatt:1952ije} one finds \eqref{eq:svIRIA}, but with 
\begin{equation}\label{eq:scalar}
    g_{L,\text{eff}}^2=g^2\,\frac{9}{2[(2L+1)!!]^2}
\end{equation}
instead of \eqref{eq:geff_QED}.
In the following, we express our results in terms of $\alpha_{\text{BSF}}$, defined as
\begin{equation}
    \alpha_{\text{BSF}} \equiv \frac{g_{L,\text{eff}}^2}{4\pi},
\end{equation}
in terms of which the capture cross section reads\footnote{
We emphasize that in the following we will always consider partial-wave cross sections with \emph{fixed} multipole order $L$, and hence the superscript $L$ will be omitted for brevity.}
\begin{equation}
    (\sigma v)_{p \lp \to n \el}^{(L)} = 4\pi \alpha_{\text{BSF}} \frac{4\omega^{2L+1}}{9} \times I_A \times I_R.
\end{equation}

\section{Bound-state formation cross section}
\label{sec:Overlap}

The central ingredient in the radiative capture cross section~\eqref{eq:svIRIA} is the radial overlap integral~\eqref{def:IR}. In this section, we provide explicit analytic expressions  for \eqref{def:IR} for transitions with multipole order $L=0,1,2$. The results are well suited for efficient numerical evaluation for arbitrary $n$, $\el$ and $\kappa=\alpha_s/\alpha_b$ and allow for the study of capture into highly excited bound states in theories with different potential strength for the initial and final state. This generalizes and systematizes known results for general $n$ for monopole~\cite{Oncala:2019yvj},  dipole~\cite{Garny:2021qsr,Biondini:2023zcz} and quadrupole~\cite{Biondini:2021ycj,Bollig:2024ipe} transitions.
Further results on the general radial overlap integral in Coulombic bound-state formation (BSF) processes are provided in Appendix~\ref{app:Radial}.

We use the shorthand definition 
\begin{align}
\label{eq:F+}
F_+(X) \,
&\equiv\,
{_2F_1}\lr -n+\el+1 ,\, X+\el+\I\zs ,\, 2\el+2 ,\, \frac{-4\I\zbb}{(\zbb-\I)^2} \rr
\end{align}
to express hypergeometric functions in our final results. 
Remarkably, we find that they can always be expressed in terms of only the single hypergeometric function $F_+(0)$.
The squared overlap integral can then be expressed as 
\begin{equation}\label{eq:IR}
\begin{aligned}
I_R\,=\,& 
\frac{2^{4 \el+2} \zbb^{2 \el+3} }{ (\mu v)^{3+2L}\lr 1+\zbb^2\rr^{2\el+4}}
\frac{\Gamma(\lp+1)^2  \Gamma (n+\el+1) }{n \Gamma (2 \el+2)^2 \Gamma (n-\el)}
S_{\lp}(\zs)\,\E^{-4\zs \gbb }
\\&\times
 \left|\frac{ 1-\E^{2\I\lr2(n-\el)\gbb-\gamma_F-\gamma_R\rr}}
{ n\kappa\zbb\lr\zbb^2- 1+\frac{2}{\kappa} \rr} \right|^2
 \left|F_+(0)\right| ^2\left|R^L_{\lp-\el}\right|^2\,,
	\end{aligned}
\end{equation}
where $\gbb$, $\gamma_F$ and $\gamma_R$ denote the phases of $(\I+\zbb)$, $F_+(0)$ and $R^L_{\lp-\el}$, respectively. Furthermore, 
\begin{equation}\label{eq:SE}
 S_{\lp}(\zs)  \equiv 
\frac{\E^{\pi\zs}\pi \zs }{\sinh(\pi\zs)} \prod_{j=1}^{\lp} \lr 1+\zs^2/j^2\rr 
=
\E^{\pi\zs}\frac{\left| \Gamma\lr 1+\lp-\I \zs \rr \right|^2}{\Gamma(1+\lp)^2}
\end{equation}
is a combination that is well-known from Sommerfeld factors, and $R^L_{\lp-\el}$
are rational functions of $n$, $\ell$, $\zeta_n$ and $\zeta_s$. 
For $L=0,1,2$, we find 

\begin{align}
R_0^0 \equiv\,& n\zbb(1-\kappa)(1+\zbb^2)(1+\el-\I\zs),
\\[5pt]
R_1^1 \equiv \,& \,
\zs(1+\zbb^2) +n\zbb(1-\kappa)\lr 2+2\I n\zbb(1-\kappa)+(\el+1)(1+\zbb^2)  \rr 
,
\\[5pt] 
R_{-1}^1 \equiv \,& 
\left[
\zs(1+\zbb^2) +n\zbb(1-\kappa)\lr 2 +2\I n\zbb(1-\kappa)-\el(1+\zbb^2) \rr 
\right]
\lr \el-\I\zs\rr \lr1+\el-\I\zs \rr
,
\\[5pt]
\notag
\label{eq:R2}
R_2^2 \equiv \,& 
\frac{n \zbb}{(\zs^2+(\el+2)^2)(1+\zbb^2)}
\left[ 
(1+\zbb^2)^2 (\el+1) (2 \el+3) (2 +\el(1-\kappa))
\right.\\*&
\notag 
+2 (1+\zbb^2)(\I \zs+\el+2)\left(2+2\el(1-\kappa)+\I n\zbb (1-\kappa)(2\el(1-\kappa)+3) \right)
\\* &
\left.
+4 (1-\kappa) (\I \zs+\el+2) \, (1+\I n \zbb (1-\kappa)) \, (2+\I n \zbb (1-\kappa)) \right]
,
\\[5pt]
\notag
\label{eq:R-2}
R_{-2}^2\equiv \,&
\frac{(\el-\I \zs) (1+\el-\I \zs)}{1+\zbb^2} \Big[
\left(1+\zbb^2\right)^2 \el (2 \el-1) ((1-\el) n \zbb  (1-\kappa)+2 \zs)
\\*&
\notag
+2 \left(1+\zbb^2\right) (1-\el+\I \zs) (2\zs -  n\zbb(1-\kappa) (\I (2 \el-1) n \zbb  (1-\kappa)-\I\zs+2 (\el-\I\zs)))
\\*&
-4 n \zbb (1-\kappa ) (1-\el+\I \zs) (-2+\I n \zbb (1-\kappa ) (-3 -\I n \zbb (1-\kappa )))
\Big]\,,
\\
\notag
R^2_0 \equiv \, &
\frac{2n \zbb\lr1+\el-\I\zs\rr}{1+\zbb^2}  \biggl[ 
(1+\zbb^2)\lr
    2-3 (1-\kappa)(1-\I \zs)  \rr
 \\*& 
+
2(1-\kappa)\lr \I n\zbb(1-\kappa)(\I n\zbb(1-\kappa)+3)+2\rr
\biggr].
\end{align}
We checked agreement with~\cite{Garny:2021qsr,Binder:2023ckj} for $L=1$ ($\lp=\el\pm1$) up to $n=3$ for all possible $\el\leq n-1$ analytically, and up to $n= 10^3$ numerically. In addition, our result for $L=1$ reduces to the expression given in~\cite{Biondini:2023zcz} in the special case $\kappa=1$. The radial overlap integral for $L=2$ was also checked 
against the quadrupole contribution in~\cite{Biondini:2021ycj} for the values $\kappa=1$, $\lp=0,2$ and $n\leq 3$, $\el=0$. For $L=0$ (thus $\el=\lp$) our result can
equivalently be written in terms of $F_+(1)$ (see Appendix~\ref{app:Radial}), yielding
\begin{equation}\label{eq:IR00 explicit}
I_R\big|_{L=0} =
\frac{2^{4 \el+2} \zbb^{2 \el+5} }{ (\mu v)^3\lr 1+\zbb^2\rr^{2\el+4}}\E^{-4\zs \gbb } S_{\el}(\zs)(\el!)^2
\frac{  \Gamma (n+\el+1) }{n \Gamma (2 \el+2)^2 \Gamma (n-\el)}
(1-\kappa)^24n^2
 \left|F_+(1)\right| ^2 \,.
\end{equation}
In this form, agreement with~\cite{Oncala:2019yvj} can be checked for any $n$ and $\el$. We remark that this rewriting in terms of $F_+(1)$ is possible for $L=0$ but does not apply generally. 
Note that for $L=0$ all transition rates vanish for $\kappa=1$. This follows from $Q_{00,fi} = \langle n\el m|\vec{p}\rangle$ and  the orthogonality of scattering- and bound-state wave functions when the potentials in the initial and final state are identical. 

\begin{figure}[!t]
\begin{center}
 \includegraphics[width=0.72\textwidth]{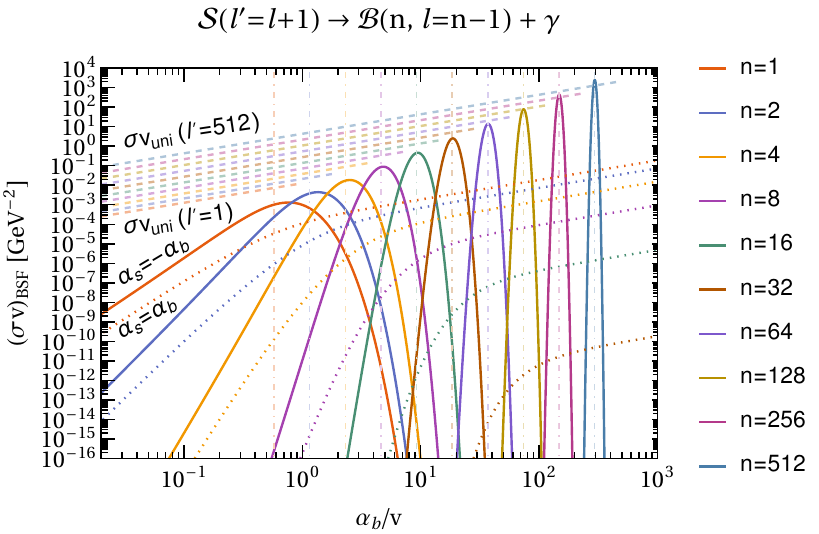}\\[0.4cm]
 \includegraphics[width=0.72\textwidth]{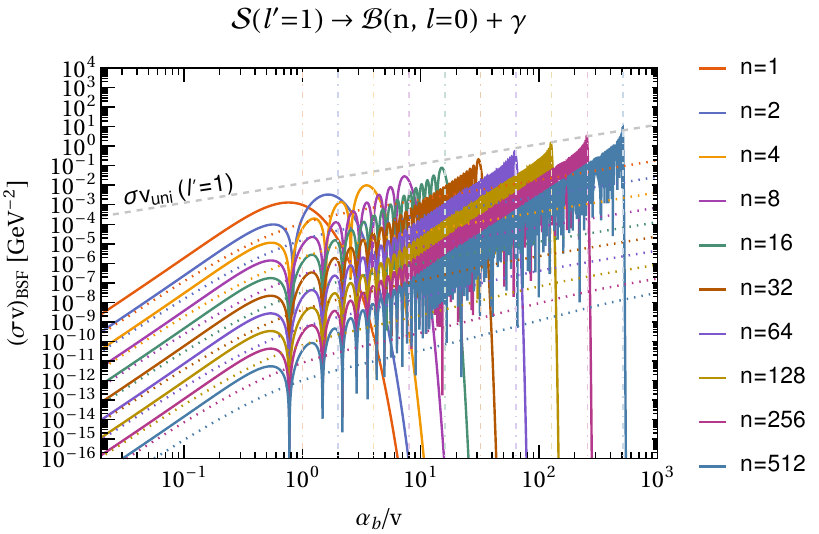}
\end{center}
\caption{\label{fig:sigvBSF} \small
 Dependence of the BSF cross section $ (\sigma v)_{p \lp \to n \el}$ on $\alpha_b/v$ for various $n$ and maximal $\ell=n-1$ (upper panel) resp. minimal $\ell=0$ (lower panel). Solid lines correspond to $\alpha_s/\alpha_b=-1$, while dotted lines show $\alpha_s/\alpha_b=1$ for comparison. Dashed lines indicate the perturbative partial-wave unitarity limit, for $\ell'=n$ (various lines in upper panel) and $\ell'=1$ (lower panel). We use 
 $L=1$, $\alpha_{\rm BSF}=0.1$, 
 $\alpha_b=0.3$ and $M=100$\,GeV. Vertical lines show the peak velocities $v=\sqrt{3}\alpha_b/n$ (upper panel) and $v=\alpha_b/n$ (lower panel).
}
\end{figure}

We briefly summarize some general properties of the dependence of the capture cross section $ (\sigma v)_{p \lp \to n \el}$ on the relative velocity $v$ of the initial state. This dependence is illustrated in Fig.~\ref{fig:sigvBSF}, for the dipole transitions with $\lp=\el+1$ at maximal $\el=n-1$ (upper panel) and minimal $\el=0$ (lower panel), for various values of $n$.\footnote{For further analysis and interpretation we refer to Sec.~\ref{sec:UV} where the special case $\el=n-1$ is investigated more closely.} 
Here we discuss some common features:
\begin{itemize}
\item In the perturbative limit $v\gg \alpha_b,|\alpha_s|$, i.e.~for $|\zeta_s|,n\zeta_n\ll 1$, BSF is suppressed as a power-law $ (\sigma v)_{p \lp \to n \el} \propto (1/v)^{2\el+6-2L}$.
This scaling arises from the dependence on $k^{2L+1}\propto v^{4L+2}$ in \eqref{eq:svIRIA},
the prefactor $\zeta_n^{2\el+3}/(\mu v)^{2L+3}\propto (1/v)^{2\el+2L+6}$ of $I_R$ in \eqref{eq:IR}, as well
as $R^L_{\lp-\el}\propto 1/v$ and $F_+(0)\to 1, S_{\lp}(\eta_s)\to 1, e^{-4\zeta_s\gamma_n}\to 1$ in this limit. Furthermore, the first factor in the second line of \eqref{eq:IR} becomes independent of $v$.
\item In the low-velocity regime $v\ll \ab/n,|\as|$, i.e.~for $|\zs|\gg 1$, $\zbb\gg 1$, the behaviour depends crucially on the sign of $\as$. For an attractive scattering potential $\as>0$, the asymptotic scaling  $ (\sigma v)_{p \lp \to n \el}\propto 1/v$ coincides with the one of  the perturbative unitarity cross section, and $(\sigma v)_{p \lp \to n \el}$
stays below $(\sigma v)^{\rm uni}_{\ell'}$ for sufficiently small coupling.  On the other hand, for a repulsive scattering potential $\as<0$, an exponential suppression occurs for $v\ll \ab/n,|\as|$, as captured already by the s-wave Sommerfeld factor $S_{0}(\zs)$ in~\eqref{eq:SE}.
\item For intermediate velocities, the BSF cross section interpolates between the low- and high-velocity limits described above. For the most familiar case $\alpha_s=\alpha_b$, we find numerically a smooth behaviour between the power-law scalings $ (\sigma v)_{p \lp \to n \el} \propto (1/v)^{2\el+6-2L}$ and $ (\sigma v)_{p \lp \to n \el} \propto 1/v$ at $v$ above and below $\alpha_b$, respectively. In contrast, for $\alpha_s<0$, the BSF cross section while being exponentially suppressed at very low velocities, can still be rather sizeable for intermediate velocities. As observed before~\cite{Oncala:2019yvj,Garny:2021qsr,Binder:2023ckj}, for $\el=n-1$ it features a pronounced ``peak'' shape as a function of velocity, with a maximum when $\zeta_n$ is of order unity, i.e. when the relative momentum $p=\mu v$ of the scattering state is of order of the Bohr momentum $p_n=\alpha_b\mu/n$. This maximum plays an important role in the context of perturbative unitarity violation~\cite{Binder:2023ckj}, as will be shown in detail below. In addition, for $\el<n-1$ the BSF cross section can have multiple peaks and oscillatory features within the intermediate velocity regime. Specifically, for $\alpha_s<0$ and $\lp>\el$, the number of oscillations is related to the number of nodes $n-\el-1$ of the radial bound-state wave function. In~\eqref{eq:IR} the strongly oscillatory behaviour that occurs for large values of $n-\el$ is encapsulated in the first factor in the second line,\footnote{
The factor is ensured to remain regular for $\zbb^2=1-2/\kappa$, see discussion at the end of Appendix~\ref{app:2F1}.} 
where the phase $\E^{4\I (n-\el)\gamma_n}$ oscillates rapidly in $v$.
\end{itemize}
These general features can be recognized in Fig.~\ref{fig:sigvBSF}. In particular, the peak structure of the BSF cross section for $\as/\ab=-1$ (solid lines) is clearly visible, with a single peak for maximal $\el$, and $n$ peaks for minimal $\el=0$. The absolute maximum occurs when the Bohr momentum is of the order of the relative momentum of the scattering state, as indicated by the vertical dot-dashed lines for the various bound state levels $n$.\footnote{The precise formula of the depicted vertical lines is derived in  \eqref{eq:zetapeak}.} The BSF cross sections close to the maxima can exceed the perturbative partial-wave unitarity bounds for the corresponding values of $\lp$ (dashed lines), as will be discussed in detail below. For comparison, the QED case $\as/\ab=1$ is shown (dotted lines), for which the smooth transition between the low- and high-velocity regimes is clearly evident and unitarity violation is absent.

\section{Enhancement of radiative bound-state formation}\label{sec:UV}

In this section we investigate under which conditions the strong enhancement of radiative bound-state formation for highly excited states occurs. In particular, we provide instructive analytical results, focusing on the simplest case of bound states with maximal angular momentum $\ell=n-1$. This also allows us to extend the analysis to transitions of arbitrary multipole order $L$.
After discussing some insights from the analytical result for the radial overlap integral we derive a large-$n$ expansion, and finally investigate whether the BSF cross section exceeds the perturbative unitarity limit.

\subsection{Analysis based on the full analytical result}

The analytical result for the BSF cross section from Sec.~\ref{sec:Overlap}
can be simplified considerably for bound states with maximal angular momentum, i.e. $\ell=n-1$, since their radial wave function ${\cal B}_{n\ell}(r)\propto r^\ell e^{-p_nr}$ is particularly simple. We further consider a scattering state with maximal angular momentum $\lp=\el+L$ for simplicity. In these cases, the radial overlap integral simplifies for {\em any} multipole order $L$ to 
\begin{equation}\label{eq:IRmaxl}
  I_R\Big|_{
 {\scriptstyle\lp=\el+L,}\atop{\scriptstyle\el=n-1~}} = \frac{2^{4n+2L}\zeta_n^{2n+3}}{(1+\zeta_n^2)^{2n+2L+2}(\mu v)^{2L+3}} S_{\ell'}(\zeta_s)e^{-4\zeta_s\arccot(\zeta_n)}\frac{(\ell'!)^2}{n(2\ell+1)!}\,(n(1-\kappa)+L)^2\,.
\end{equation}
This implies for the ratio of the BSF cross section for general $\kappa=\as/\ab$ to the abelian case $\kappa=1$,
\bea\label{eq:AbelianRatio}
 r_L\left(n,\frac{\alpha_b}{v},\kappa\right) &\equiv&
 \frac{ (\sigma v)_{p \lp \to n \el} }{  (\sigma v)_{p \lp \to n \el}  \big|_{\kappa=1}}\,\Bigg|_{
 {\scriptstyle\lp=\el+L,}\atop{\scriptstyle\el=n-1~}}
 \nn\\
 &=& \exp\left[4\frac{\alpha_b}{v}(1-\kappa)\arccot\left(\frac{\alpha_b}{nv}\right)\right] \lr 1+\frac{n}{L}(1-\kappa)\rr^2\frac{S_{\lp}(\kappa \alpha_b/v)}{S_{\lp}(\alpha_b/v)}\,,
\eea
where $S_{\ell'}$ are the Sommerfeld factors defined in~\eqref{eq:SE}.

We first discuss the particularly simple case $\kappa=-1$, for which the ratio of the Sommerfeld factors reduces to a single exponential, \cf~\eqref{eq:SE}, independently of $\lp$, giving
\begin{equation}\label{eq:AbelianRatiokappaminusone}
 r_L\!\left(n,\frac{\alpha_b}{v},-1\right) \,=\,
 \exp\left\{8\frac{\alpha_b}{v}\left[\arccot\left(\frac{\alpha_b}{nv}\right)-\frac{\pi}{4}\right]\right\} \lr 1+\frac{2n}{L}\rr^2\,.
\end{equation}
We stress that we did not have to make any approximations to arrive at this result. It exemplifies the drastic difference between the case $\as=-\ab$ compared to $\alpha_s=\alpha_b$. For $\alpha_b/v\ll 1$, the ratio approaches the constant $(1+2n/L)^2$. The argument of the exponential is positive for $0<\alpha_b/v<n$, and turns negative for $\ab/v>n$ (note that arccot$(1)=\pi/4$). Thus, the ratio of BSF cross sections is exponentially enhanced as long as the Bohr momentum $p_n=\ab\mu/n$ is smaller than the relative momentum $p=\mu v$ of the scattering state, $p_n<p$, and exponentially suppressed for $p_n>p$. This is consistent with the strongly peaked shape of the BSF cross section for $\el=n-1$ and $\kappa=-1$, compared to a smooth dependence on $\ab/v$ for $\kappa=1$, as observed in the upper panel of Fig.~\ref{fig:sigvBSF}. Employing $\zeta_n=\alpha_b/( n v)$, the maximal enhancement can be obtained by writing~\eqref{eq:AbelianRatiokappaminusone} as
\be\label{eq:AbelianRatiokappaminusone2}
  r_L\!\left(n,\frac{\alpha_b}{v},-1\right) \,=\, \lr 1+\frac{2n}{L}\rr^2 e^{n f(\zeta_n)}\,,
\ee
where $f(\zeta_n)\equiv 8\zeta_n \left(\arccot\left(\zeta_n\right)-\frac{\pi}{4}\right)$. This function has a maximum at $\zeta_n^*\approx 0.4416$ given by $f_*\equiv f(\zeta_n^*)\approx 1.306$. 

Let us now return to the general case $\kappa\not= -1$. The qualitative behaviour of $r_L$ is similar to the one for $\kappa=-1$ as long as $\kappa<0$, i.e. for a repulsive scattering potential.
In particular, the first factor in~\eqref{eq:AbelianRatio} yields an exponential enhancement, since $\alpha_b>0$ and $\pi/2\geq \arccot(\zeta_n)>0$ for $0\leq\zeta_n<\infty$  (note that this only requires $\kappa<1$). The argument of the exponential in~\eqref{eq:AbelianRatio} approaches zero both in the limits $\ab/v\ll 1$ as well as for $\ab/v\gg n$. The second factor  in~\eqref{eq:AbelianRatio} corresponds to a polynomial enhancement for large~$n$. The third factor in~\eqref{eq:AbelianRatio} approaches unity for $\ab/v\to 0$, and yields an exponential suppression for $\ab/v\gg 1$ if $\kappa<0$. This suppression ultimately dominates over the first factor for very low velocities $\alpha_b/v\gg n$. Altogether, this yields a strongly peaked shape of the BSF cross section for $\el=n-1$ and any $\kappa<0$, similar to the one shown in the upper panel of Fig.~\ref{fig:sigvBSF} for $\kappa=-1$. The precise relative velocity for which the ratio is maximal depends on $\kappa$, but lies within the regime for which $p_n$ is of order $p$, i.e. occurs for $\zbb$ of order unity. More concrete analytical results can be obtained by considering the large-$n$ limit, to which we turn next.

\subsection{Large-\emph{n} expansion}

The strong enhancement of the BSF cross section for repulsive scattering states and large~$n$ raises the question whether it respects the perturbative unitarity bound~\eqref{eq:unitaritybound}. Following the numerical analysis in~\cite{Binder:2023ckj}, we investigate this question analytically here.
For that aim, we focus on the particularly simple BSF process discussed above, with maximal $\ell=n-1$ and $\ell'=\ell+L$. Furthermore, following
the previous discussion, we take the large-$n$ limit while choosing $v$ such that $\zeta_n=\alpha_b/(nv)$ remains fixed. We also keep $\alpha_s$ and $\alpha_b$ fixed when taking the limit, and assume either (i) that $\kappa=\alpha_s/\alpha_b$ is different from unity by an amount at least of order $1/n$, or (ii) that $\kappa=1$ precisely. Using the analytical result ~\eqref{eq:IRmaxl} that applies to radiative BSF transitions of any multipole order $L\geq 0$, we expand the ratio of the BSF cross section~\eqref{eq:svIRIA}, $ (\sigma v)_{p \lp \to n \el}$ to the corresponding maximal cross section $(\sigma v)_{\lp}^\text{uni}$ allowed by perturbative unitarity~\eqref{eq:unitaritybound}. Using  Stirling's formula for expanding the Gamma function for large (complex, see~\eqref{eq:SE}) argument,
we find 
\begin{align}\label{eq:UVSeries}
  \ln  \left\lbrace \frac{ (\sigma v)_{p \lp \to n \el} }{(\sigma  v)_{\lp}^\text{uni}}\right\rbrace  \Bigg|_{
  {\scriptstyle\lp=\el+L,}\atop{\scriptstyle\el=n-1~}}
  = \,n\, P(\zbb,\kappa)\,+\ln\lr n \rr\, Q(\zeta_n,\kappa) +\mathcal{O}(n^0).
\end{align}
The leading term is proportional to $n$ with coefficient
\begin{equation}\label{eq:UV NLO}
  P(\zbb,\kappa) \equiv 2\zbb\kappa\lr\frac\pi2-\text{arctan}(\zbb\kappa)-2\arccot(\zbb)\rr+\ln\lr\frac{4\zbb^2(1+\zbb^2\kappa^2)}{(1+\zbb^2)^2}\rr ,
\end{equation}
and the logarithmic correction is given by
\be\label{eq:Q}
   Q(\zeta_n,\kappa) = \left\{\begin{array}{ll}
    \phantom{-}\frac12 & \kappa\not=1 \,,\\[0.2cm]
    -\frac32 & \kappa=1 \,.
  \end{array}\right.
\ee
We note that $Q(\zeta_n,\kappa)$ is in fact independent of $\zbb$.
From \eqref{eq:UV NLO}, \eqref{eq:Q} we recover the exponential enhancement of the ratio of BSF cross sections for $\kappa=-1$ compared to $\kappa=1$ derived in~\eqref{eq:AbelianRatiokappaminusone2} above, from
$P(\zeta_n,-1)-P(\zeta_n,1)=\zeta_n(8\,\arccot(\zeta_n)-2\pi)=f(\zeta_n)$. Furthermore, $Q(\zeta_n,-1)-Q(\zeta_n,1)=2$ accounts for the additional power-law enhancement $\propto n^2$ obtained from expanding the prefactor of the exponential on the right-hand side of~\eqref{eq:AbelianRatiokappaminusone2} for large~$n$.

\begin{figure}[!t]
\begin{center}
 \includegraphics[width=0.75\textwidth]{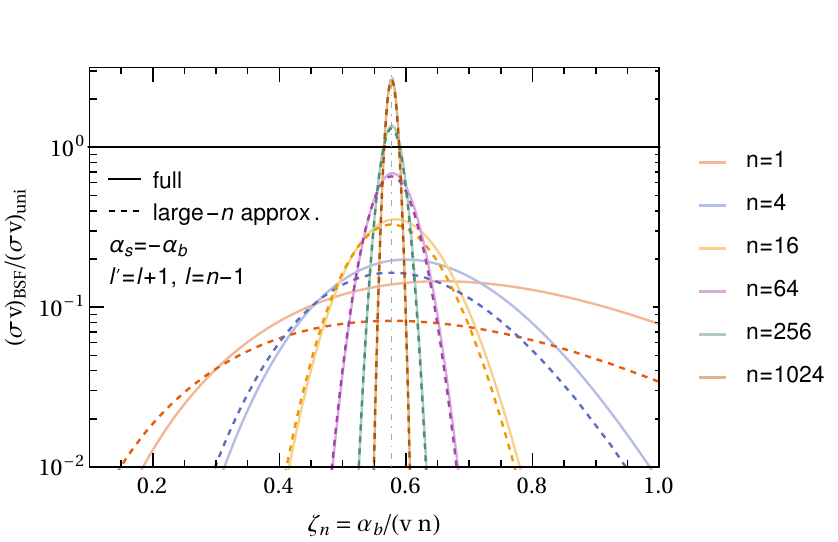}
\end{center}
\caption{\label{fig:sigvBSFlargenapprox} \small
 Comparison of the full quantum-mechanical result for the BSF cross section (solid lines) to the large-$n$ approximation~\eqref{eq:UVSeries} (dashed lines) for various values of $n$, and $\el=n-1, \lp=\el+1$. The cross sections are normalized to the corresponding perturbative unitarity bound. We used $L=1, \alpha_\BSF=0.1, \ab=-\as=0.3$ ($\kappa=-1$).  
 The constant ($n$-independent) term in~\eqref{eq:UVSeries} was adjusted to ensure an overall normalization factor that agrees with the full result for the largest~$n$. 
 The peak position $\zbb^\text{peak}=1/\sqrt{1-2\kappa}=1/\sqrt{3}\approx 0.577$ is indicated by the dot-dashed vertical line.
}
\end{figure}

Next, we compare the large-$n$ expansion~\eqref{eq:UVSeries} to the full quantum-mechanical result. For the case $\as=-\ab$ shown in Fig.~\ref{fig:sigvBSFlargenapprox}, we observe that~\eqref{eq:UVSeries} provides a good approximation even for moderate values of~$n$. As expected, it also captures the relevant features, i.e. the shape of the peak of the BSF cross section as a function of velocity (or equivalently~$\zbb$) and the enhancement of the peak value for large $n$. 

\begin{figure}[!t]
\begin{center}
 \includegraphics[width=0.75\textwidth]{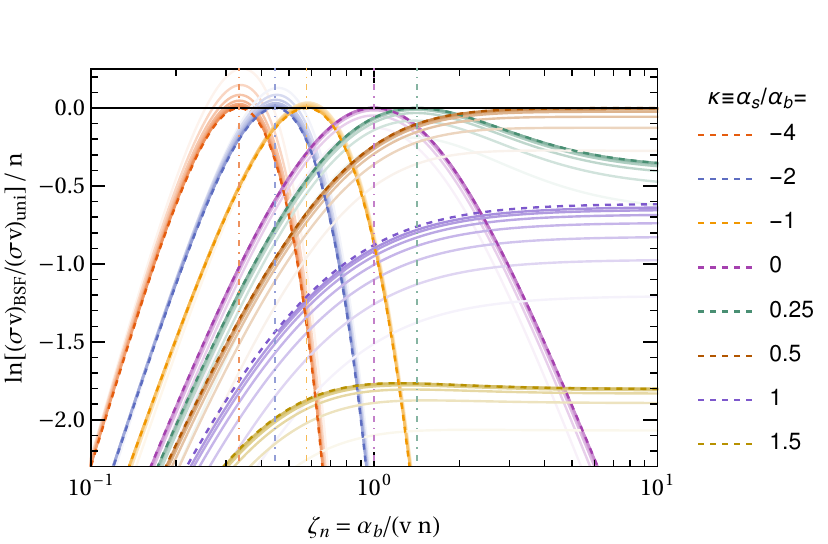}
\end{center}
\caption{\label{fig:sigvBSFlargenapproxKappa} \small
Convergence of the BSF cross section $ (\sigma v)_{p \lp \to n \el}$ towards the large-$n$ approximation~\eqref{eq:UVSeries} for $\el=n-1$, $\lp=\el+L$, and various values of $\kappa$. Dashed lines show the dependence of the leading large-$n$ result $P(\zbb,\kappa)$ on $\zbb$. For each value of $\kappa$, the solid line shows $\ln [ (\sigma v)_{p \lp \to n \el} / (\sigma  v)_{\lp}^\text{uni}]/n$ for $n=8,16,32,\dots,512$, with increasingly darker colours for increasing~$n$. For $\kappa<1/2$, $P(\zeta_n,\kappa)$ has a maximum at $\zbb^\text{peak} = 1/\sqrt{1-2\kappa}$ (vertical lines). We used $L=1, \alpha_\text{BSF}=0.5, \ab=0.3$.  
}
\end{figure}

The large-$n$ expansion allows us to analytically derive various results concerning the behaviour of the BSF cross section.
In particular, for $\zbb$ of order unity and for large~$n$, the dependence of the BSF cross section on the relative velocity is determined by the dependence of $P(\zbb,\kappa)$ on $\zbb$.
Using~\eqref{eq:UV NLO}, we observe a qualitatively different behaviour depending on the value of $\kappa$:
\begin{itemize}
  \item[(1)] 
  For $\kappa<1/2$, $P(\zeta_n,\kappa)$ has a maximum at
  \be\label{eq:zetapeak}
    \zeta_n^\text{peak} = \frac{1}{\sqrt{1-2\kappa}}\,,
  \ee
  leading to a peak\footnote{We note that for $\kappa=-1$, $\zeta_n^\text{peak}=1/\sqrt{3}\approx 0.577$ is distinct from the value $\zeta_n^*$ at which the function $f(\zeta_n)$ defined below~\eqref{eq:AbelianRatiokappaminusone2} is maximal. The latter is related to the {\it ratio} of cross sections~\eqref{eq:AbelianRatiokappaminusone}, while the former is related to the large-$n$ limit of the cross section itself.} of the BSF cross section as function of relative velocity, as observed above. This result implies that for large $n$ the peak occurs at
  \be\label{eq:vpeak}
    v_\text{peak}(n,\kappa)=\alpha_b/n \times(\zeta_n^\text{peak})^{-1} = \sqrt{1-2\kappa}\,\alpha_b/n\,.
  \ee
  The peak positions according to this result are shown as vertical dot-dashed lines in the upper panel of Fig.~\ref{fig:sigvBSF}, and agree well with the peaks of the full BSF cross section for large $n$. This result also explains why the peaks are approximately on top of each other when using $\zeta_n$ as variable, such as in Fig.~\ref{fig:sigvBSFlargenapprox}.

  We emphasize  that the maximum of $P(\zeta_n,\kappa)$ is simultaneously the (only) root of the function and its value is thus given by 
  \be
    P(\zeta_n^\text{peak},\kappa) = 0\,,
  \ee
  while $P(\zeta_n,\kappa) < 0$ for $\zeta_n \not= \zeta_n^\text{peak}$.
  This has important implications for $\UV$ that we discuss below. 
  
  \item[(2)] For $\kappa>1/2$, the function $P(\zbb,\kappa)<0$ is always negative. 
\end{itemize}
The dependence of $P(\zeta_n,\kappa)$ on $\kappa$ is illustrated by the dashed lines in Fig.~\ref{fig:sigvBSFlargenapproxKappa}. For each $\kappa<1/2$, the position of the maximum~\eqref{eq:zetapeak} is indicated by a vertical line. The behaviour of $P(\zeta_n,\kappa)$ for large $\zeta_n$ can also be understood analytically:
\begin{itemize}
  \item[(1)] For $\kappa<0$, expanding~\eqref{eq:UV NLO} for large $\zeta_n$ yields $P(\zeta_n,\kappa)=2\pi\kappa\zeta_n+{\cal O}(\zeta_n)^0$. The linear dependence on $\zeta_n$ with a negative
  prefactor corresponds to the exponential suppression of the BSF cross section in the limit $v\to 0$ due to the repulsive scattering state, as discussed above, which arises from the Sommerfeld factor~\eqref{eq:SE}. In Fig.~\ref{fig:sigvBSFlargenapproxKappa} this corresponds to the decline at large $\zeta_n$, for all curves with $\kappa<1/2$.\footnote{Note that the vertical axis in Fig.~\ref{fig:sigvBSFlargenapproxKappa} corresponds to a linear scale for $P(\zbb,\kappa)$, while the horizontal axis is logarithmic in $\zbb$.}
  \item[(2)] For $\kappa>0$, i.e. for an attractive scattering state, 
  \be
    P(\zeta_n,\kappa) = 2(1-2\kappa)+2\ln(2\kappa) + {\cal O}(\zeta_n)^{-1}\,,
  \ee
  approaches a constant value for large $\zeta_n$. This constant value is strictly negative for $0<\kappa<1/2$ as well as for $\kappa>1/2$, and exactly zero only for $\kappa=1/2$.
\end{itemize}
The qualitative differences in the dependence of $P(\zeta_n,\kappa)$ on $\zeta_n$ for the cases $\kappa<0, 0<\kappa<1/2$ and $\kappa>1/2$, respectively, can be recognized by the dashed lines in Fig.~\ref{fig:sigvBSFlargenapproxKappa}.

In summary, the function $P(\zeta_n,\kappa)\leq 0$ is non-positive within the entire physically sensible parameter regime ($\zeta_n>0$, real-valued $\kappa$). It can be $0$ only for $\kappa<1/2$, and takes this value at $\zeta_n=\zeta_n^\text{peak}$.

\subsection{Perturbative unitarity violation for $\el=n-1$}\label{sec:puvmaximumell}

Here we show analytically under which conditions the BSF cross section obtained from the usual quantum-mechanical transition matrix element can saturate or even exceed the upper bound~\eqref{eq:unitaritybound} related to perturbative unitarity. For simplicity we still focus on the single BSF process discussed above, i.e. a bound state with maximal $\el=n-1$ and a scattering state with $\lp=\el+L$. The multipole order $L$ of the transition is kept as a free parameter, as well as the principal quantum number $n$.

We first consider the $n$-dependence of this BSF cross section, keeping $\zbb=\ab/(n v)$ (and $\kappa=\alpha_s/\alpha_b$) fixed. In the large-$n$ limit~\eqref{eq:UVSeries}, the ratio of the BSF cross section and the unitarity bound depends crucially on the sign of $P(\zbb,\kappa)$ defined in~\eqref{eq:UV NLO}. As shown above, $P(\zbb,\kappa)\leq 0$. This implies that for large enough $n$, this cross section is guaranteed to be {\it below} the upper limit~\eqref{eq:unitaritybound}.  However, this is only true when $P(\zbb,\kappa)$ is {\it strictly} negative. As derived above, $P(\zbb,\kappa)=0$ occurs only for $\kappa<1/2$ and $\zbb=\zbbpeak$, i.e. at the relative velocity $v=v_\text{peak}(n,\kappa)$ (see~\eqref{eq:vpeak}) for which the BSF cross section attains its maximum in the large-$n$ limit. In this case, the next term in the large-$n$ expansion~\eqref{eq:UVSeries} becomes relevant. Using~\eqref{eq:Q}, this implies that
\be\label{eq:puvformaximumell}
   \frac{ (\sigma v)_{p \lp \to n \el}\big|_{v=v_\text{peak}(n,\kappa)} }{(\sigma  v)_{\lp}^\text{uni}}\Bigg|_{{\scriptstyle\lp=\el+L,}\atop{\scriptstyle\el=n-1~}}\!
  = n^{1/2} \times c_L(\kappa) \,\left(1+{\cal O}(n^{-1})\right)\,,
\ee
with some coefficient $c_L(\kappa)$. For $\kappa>1/2$ or $v\not= v_\text{peak}(n,\kappa)$, the ratio is instead suppressed exponentially in the large-$n$ limit. 
This behaviour is reflected in the example of $L=1$ and $\kappa=-1$ shown in Fig.~\ref{fig:sigvBSFlargenapprox}. As $n$ increases, the dependence of the BSF cross section on $\zeta_n$ becomes more and more peaked. We verified numerically that the increase of the peak height  in Fig.~\ref{fig:sigvBSFlargenapprox} is consistent with the $n^{1/2}$-scaling of \eqref{eq:puvformaximumell}.

For large but finite values of $n$, the unitarity bound can therefore be violated in a narrow region around the peak of the BSF cross section as can be seen for the numerical examples shown in Figs.~\ref{fig:sigvBSFlargenapprox} and ~\ref{fig:sigvBSFlargenapproxKappa}. 

\begin{figure}[t]
\begin{center}
\includegraphics[width=.58\textwidth]{"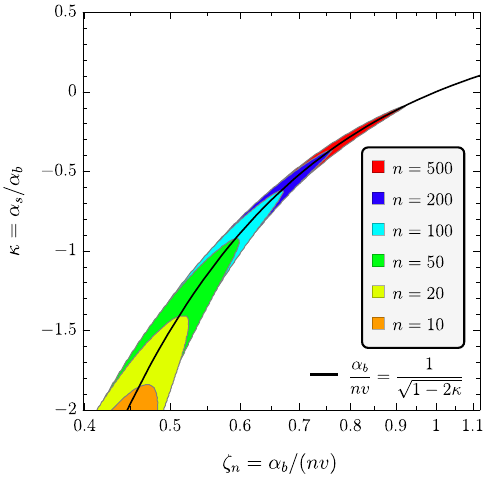"}
\caption{\label{fig:kappazbb}
Region in parameter space spanned by $\zbb=\ab/(n v)$ versus $\kappa=\as/\ab$ for which the quantum-mechanical BSF cross section $ (\sigma v)_{p \lp \to n \el}$ (for $\el=n-1,\,\lp=\el+L$) exceeds the perturbative unitarity bound~\eqref{eq:unitaritybound}. In the large-$n$ limit this happens only along the black dashed line  $\zbb=\zbbpeak\equiv 1/\sqrt{1-2\kappa}$. For large but finite $n$, $\UV$ occurs in a narrow strip around this line (coloured regions), the width of which depends on the chosen parameters. We used $L=1,\, \alpha_\BSF=0.1$, $\ab=0.4$. }
\end{center}
\end{figure}

To illustrate this result, the  parameter space in $\zbb$ and $\kappa$ for which the BSF cross section violates the unitarity limit is shown in Fig.~\ref{fig:kappazbb}. In the limit $n\to \infty$, this occurs only on the black line which corresponds to $\zbb=\zbbpeak\equiv 1/\sqrt{1-2\kappa}$. For finite $n$, the cross section is in excess of the unitarity limit in a narrow strip around this line.
We highlight the following points:
\begin{itemize}
\item The apparent perturbative violation of unitarity occurs even for (in principle arbitrarily) weak transition strength. The effective coupling $\alpha_\text{BSF}$ responsible for BSF enters the prefactor $c_L(\kappa)$ in~\eqref{eq:puvformaximumell} linearly. Even if $c_L(\kappa)$ is arbitrarily small, the scaling with $n^{1/2}$ leads to a BSF cross section that exceeds the unitarity limit for large enough $n$.
\item The first two coefficients in the large-$n$ expansion in~\eqref{eq:UV NLO} are independent of the multipole order $L$. This implies that also the conditions under which PUV can occur in the large-$n$ limit are valid for transitions characterized by any $L$.
\item The numerical analysis of~\cite{Binder:2023ckj} reported PUV for the BSF cross sections summed over all possible final-state bound states. Here, we find a stronger result: PUV for the capture cross section into a {\it single} bound state level  $n\ell$ (summed only over $m=-\ell,\dots,\ell$). 
\end{itemize}
The very peculiar conditions under which perturbative unitarity violation can occur call for an explanation. We offer some first steps in this direction in the following section, allowing us to perform a more general analysis covering the case $\ell<n-1$ as well as the total BSF cross section summed over all $n$ in Sec.~\ref{sec:puv}.

\section{Semi-classical approximation}
\label{sec:class}

In this section we provide an interpretation for the ``anomalous'' enhancement of the BSF cross section that can occur when the potential of the initial state differs from that of the final state in terms of a semi-classical approximation for large $\ell$ and $n$. This also allows us to understand the particular conditions under which this enhancement occurs.

Let us start with some generalities. For a central potential $V(r)$, we decompose the contribution to the wave function with  angular momentum numbers $\ell, m$ as $\psi({\bm r})=R_\ell(r)Y_{\ell m}(\hat{\vec r})$. The reduced radial wave function $u_\ell(r)=rR_\ell(r)$ satisfies 
\be
  -\frac{1}{2\mu}u_\ell''(r)=(E-V_\text{eff}(r))u_\ell(r)\,,
\ee
with effective potential
\be
  V_\text{eff}(r)=\frac{\ell(\ell+1)}{2\mu r^2}+V(r)\,.
\ee
The reduced radial wave functions $u_{s,\ell'}(r)$ and $u_{b,\ell}(r)$ for the initial scattering and final bound state are solutions for the effective potentials
\be\label{eq:Veff}
  V_{\text{eff},s}(r)=\frac{\ell'(\ell'+1)}{2\mu r^2}+V_s(r),\quad
  V_{\text{eff},b}(r)=\frac{\ell(\ell+1)}{2\mu r^2}+V_b(r)\,,
\ee
respectively, with energies $E_s=\mu v^2/2$ and $E_b=E_n=-\alpha_b^2\mu/(2n^2)$, and 
$V_{s,b}(r)$ given in \eqref{eq:HbVb}, \eqref{eq:Vs}.
The radial overlap integral~\eqref{def:IR} now reads 
\be
\label{eq:IRsemiclass}
  I_R = \left| \int_0^\infty dr\, u_{b,\ell}^*(r) r^L u_{s,\ell'}(r) \right|^2\,,
\ee
and determines the properties of the BSF cross section we are interested in.
In the following we consider $u_{s,\ell'}(r)$ and $u_{b,\ell}(r)$ in the semi-classical limit $\ell',\ell,n\gg 1$ and
develop an interpretation of the conditions under which $I_R$ can be parametrically enhanced.\footnote{We define $u_{s,\ell}(r)$ for the scattering state via
$\psi_{{\bm p}}(\bm r)=4\pi\sum_{\ell,m}\frac{u_{s,\ell}(r)}{r}Y_{\ell m}(\hat r)Y_{\ell m}^*(\hat p)$. In the following we drop the subscripts $\ell$ and $\ell'$ for brevity and denote the reduced radial wave functions simply by $u_s$ and $u_b$.}

\subsection{Formation of bound states with maximal angular momentum}\label{sec:semiclasslmax}

In Sec.~\ref{sec:UV}, we found that the cross section for the formation of bound states with maximal angular momentum $\ell=n-1$ has a dependence on the relative velocity $v$ of the initial scattering state that exhibits a single maximum, provided that $\kappa\equiv \alpha_s/\alpha_n<1/2$.
In the limit of large $n$, the maximum occurs at the velocity given in~\eqref{eq:vpeak}. Furthermore, the BSF cross section at this maximum is strongly enhanced, exceeding the unitarity bound for large enough $n$, see~\eqref{eq:puvformaximumell}.

\begin{figure}[t]
\begin{center}
\includegraphics[width=0.4\textwidth]{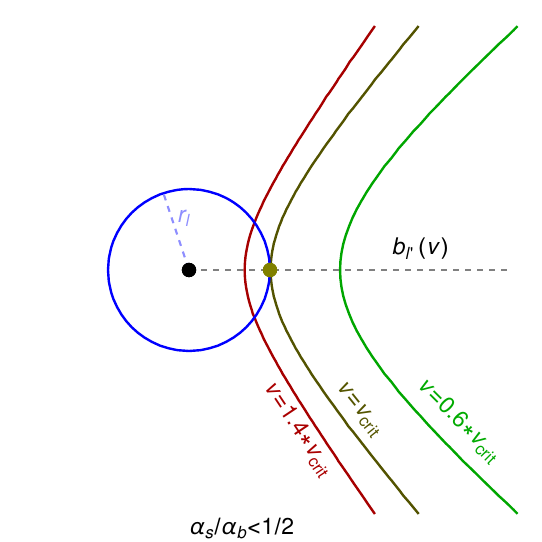}
\qquad
\includegraphics[width=0.4\textwidth]{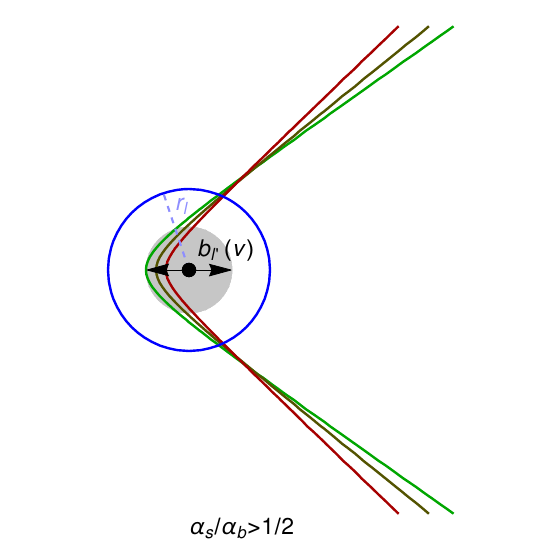}
\hfill
\caption{Classical hyperbolic orbits in the scattering-state potential $V_s(r)=-\alpha_s/r$, and circular orbit in the bound-state potential $V_b(r)=-\alpha_b/r$, that correspond to the initial and final states of the capture process with maximal $\ell=n-1$ and $|\ell'-\ell|\leq L$ in the classical limit (large $\ell,\ell',n$, fixed $L$). For $\alpha_s/\alpha_b=-0.2<1/2$ (left) the impact parameter $b_{\ell'}(v)$ matches the orbital radius $r_\ell$ of the bound state for $v=v_\text{crit}$ (see~\eqref{eq:vcrit}), corresponding to maximally enhanced BSF due to an ``anomalously'' large wave function overlap. For $\alpha_s/\alpha_b=1>1/2$ (right), the impact parameter $b_{\ell'}(v)$ is strictly \emph{smaller} than $r_\ell$ for all $v$. Figure for $\ell'=\ell+1, n=50, \ell=49$.
\label{fig:Sketch}
}
\end{center}
\end{figure}

In order to better understand the origin of this maximum, it is instructive to first consider the classical trajectories corresponding to the initial scattering and final bound state, respectively. These follow from the solutions of the classical 
equations of motion with the same angular momentum and energy as the quantum-mechanical states. For the scattering state, this corresponds to a hyperbolic orbit as shown in Fig.~\ref{fig:Sketch}. The impact parameter $b_{\lp}(v)$ corresponds to the classical turning point of the radial motion, determined by the radius $r$ for which 
\be
 V_{\text{eff},s}(r)\big|_{r=b_{\lp}(v)}=E_s\,.
\ee
Using~\eqref{eq:Veff} this gives  
\be\label{eq:impact}
  b_{\lp}(v) = \frac{1}{v\mu}\left(\sqrt{\frac{\alpha_s^2}{v^2}+\lp(\lp+1)}-\frac{\alpha_s}{v}\right).
\ee
For the bound state, the effective potential $V_{\text{eff},b}(r)$ has a minimum at radius
\be\label{eq:rell}
  r_\ell = \frac{\el(\el+1)}{\alpha_b\mu}
\ee
with energy $E=V_{\text{eff},b}(r_\el)=-\alpha_b^2\mu/(2\el(\el+1))$. The classical solution with constant radius $r_\ell$ is the circular orbit. For large $\ell\gg 1$ the circular orbit matches the energy $E_n=-\alpha_b^2\mu/(2n^2)$ and angular momentum $\el$ of the bound state with maximal $\el=n-1$, up to corrections that are relatively suppressed by $1/\el$.  

Before discussing semi-classical approximations, we present a relevant observation.
The starting point is the question under which conditions the impact parameter  can be equal to the classical circular radius,
\be\label{eq:impactmatchescircle}
  b_{\lp}(v)\big|_{v=v_\text{crit}} = r_\el\,.
\ee
This is satisfied for the ``critical'' relative velocity 
\be\label{eq:vcrit}
  v_\text{crit} = \frac{\alpha_b}{\sqrt{\ell(\ell+1)}}\,\sqrt{\frac{\ell'(\ell'+1)}{\ell(\ell+1)}-2\frac{\alpha_s}{\alpha_b}}
  = \frac{\alpha_b}{n}\sqrt{1-2\frac{\alpha_s}{\alpha_b}}\,\lr1+{\cal O}(1/n)\rr
\ee
of the scattering state, where we assumed that $|\ell'-\ell|\leq L$ remains fixed while $\ell,\ell',n$ are taken to be large in the second step.

Remarkably, this result for $v_\text{crit}$ precisely coincides with the velocity $v_\text{peak}$ from~\eqref{eq:vpeak}, for which the BSF cross section via radiative capture has its maximum.  
This yields an intuitive interpretation: the formation of a bound state is most efficient if the classical trajectories corresponding to the initial and final state ``match'' to each other, while the mismatch in energy is emitted as radiation. 
This can be recognized in the left part of Fig.~\ref{fig:Sketch}. The impact parameter $b_{\ell'}(v)$ of the hyperbolic orbit with $v=v_\text{crit}$ in the potential $V_s(r)=-\alpha_s/r$ precisely ``touches'' the circular orbit in the potential $V_b(r)=-\alpha_b/r$. 

We note that~\eqref{eq:impactmatchescircle} has a real solution only for $\kappa\equiv \alpha_s/\alpha_b\leq 1/2$ (in the limit of large $\ell,\ell',n$ considered here). For $\kappa>1/2$, the impact parameter~\eqref{eq:impact} is necessarily \emph{smaller} than the radius~\eqref{eq:rell} of the circular orbit with the same angular momentum (note that $\ell'=\ell(1+{\cal O}(1/n))$ for large $\ell',\ell,n$ at fixed $L$). Specifically,
\be
  \frac{b_{\ell'}(v)}{r_\ell}\leq \left\{
  \begin{array}{ll}
  \infty & \qquad \alpha_s<0\,,\\[0.2cm]
\displaystyle  \frac{\alpha_b}{2\alpha_s}\frac{\ell'(\ell'+1)}{\ell(\ell+1)}=\frac{1/2}{\alpha_s/\alpha_b}(1+{\cal O}(1/n)) & \qquad\alpha_s>0\,,
  \end{array}\right.
\ee
with upper bound approached for $v\to 0$.
A case with $\alpha_s/\alpha_b>1/2$ is illustrated in the right part of Fig.~\ref{fig:Sketch}, for which the impact parameter is confined to be within the grey circle inside the circular orbit of the corresponding bound state (blue). Thus, the existence of a real solution to the matching condition~\eqref{eq:impactmatchescircle}  yields an interpretation why strongly enhanced BSF occurs only for $\alpha_s/\alpha_b\leq 1/2$ (for bound states with maximal $\ell=n-1$, see below for a generalization). In particular, the matching condition can always be satisfied if the scattering potential is repulsive, $\alpha_s<0$.

\begin{figure}[t]
\begin{center}
\includegraphics[width=0.55\textwidth]{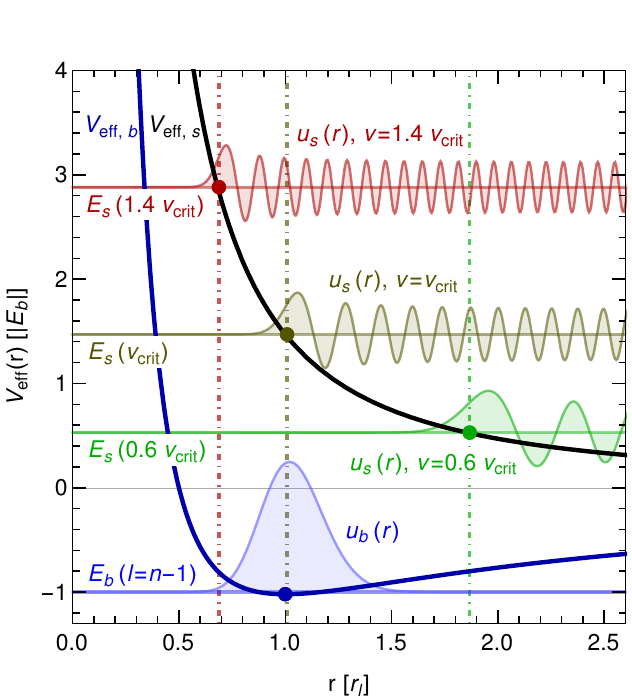}
\includegraphics[width=0.35\textwidth]{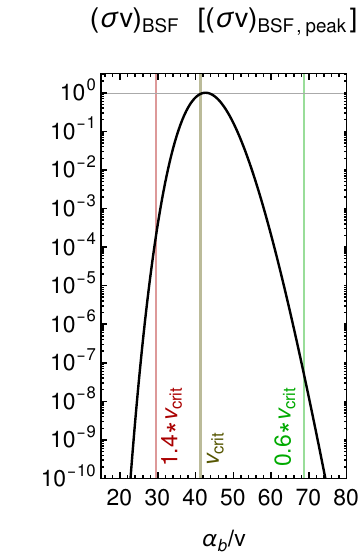}
\hfill
\caption{
\label{fig:Veff}
\emph{Left:} Effective potentials $V_{\text{eff},b}(r)$ of the bound state with angular momentum $\ell$ (thick dark blue) and $V_{\text{eff},s}(r)$ (thick black) of the scattering state with $\ell'=\ell+1$, for $\alpha_s/\alpha_b<1/2$. Also shown are the reduced wave functions $u_b(r)$ of the bound state with  $\ell=n-1$ (blue) and $u_s(r)$ of three exemplary scattering states, with relative velocities $v=1.4\,v_\text{crit}$, $v=v_\text{crit}$, and $v=0.6\,v_\text{crit}$, respectively, offset along the vertical axis to the corresponding energy eigenvalues. The impact parameters $b_{\ell'}(v)$ (i.e.~the classical turning points) are indicated by vertical dot-dashed lines. 
\emph{Right:} $ (\sigma v)_{p \lp \to n \el}$ versus $\alpha_b/v$, with vertical lines indicating the three relative velocities from the left. Plot for $n=\ell'=50, \ell=49, \alpha_b=0.9, \alpha_s/\alpha_b=-0.2$.
}
\end{center}
\end{figure}

In order to connect the purely classical picture to the quantum-mechanical result, we consider the radial wave functions $u_{b}(r)$ and $u_{s}(r)$. Their real parts are shown in the left part of Fig.~\ref{fig:Veff}, along with the corresponding effective potentials. The bound-state wave function with maximal $\ell=n-1$ is $u_{b}(r)\propto r^{\ell-1}e^{-\alpha_b\mu r/n}$. For large $\ell$ and $n$ it is peaked around the classical circular radius $r_\ell$. The scattering state wave functions $u_{s}(r)$ are exponentially suppressed within the classically forbidden region $r<b_{\ell'}(v)$, and oscillate for $r\gg b_{\ell'}(v)$. This means the radial overlap integral~\eqref{eq:IRsemiclass} entering $I_R$ becomes exponentially small when $b_{\ell'}(v)\gg r_\ell$. This case is illustrated by a scattering state with $v=0.6v_\text{crit}$ in Fig.~\ref{fig:Veff}. On the other hand, when $b_{\ell'}(v)\ll r_\ell$, the bound state is mostly supported at radii for which $u_s(r)$ oscillates, leading to a suppression of $I_R$ as well (see case $v=1.4v_\text{crit}$ in Fig.~\ref{fig:Veff}). Thus, $I_R$ becomes maximal when $v$ is close to $v_\text{crit}$.

In summary, the qualitative picture obtained here  supports the quantitative results for the bound-state formation cross section from Sec.~\ref{sec:UV}, and suggests a semi-classical interpretation. It also explains why strongly enhanced BSF does not occur in abelian theories, for which the potential strengths in the initial and final state are equal. Only for $\alpha_s/\alpha_b\leq 1/2$ can the classical trajectories of the initial and final state smoothly match onto each other, reflecting the ``anomalously'' enhanced radial overlap for initial relative velocities close to $v_\text{crit}$ within the quantum-mechanical computation. 

The validity of this qualitative picture can be further confirmed (see Appendix~\ref{sec:Airy}) by suitable semi-classical approximations for the radial wave functions, which show that the resulting radial overlap integral $I_R$ reproduces the scaling of the peak value of the BSF cross section with $n$ obtained in~\eqref{eq:puvformaximumell} from the large-$n$ approximation of the full quantum-mechanical result. Furthermore, both the qualitative and quantitative semi-classical description can be extended to general bound states with non-maximal $\ell$ for large $n$, as we discuss next.

\subsection{Formation of bound states  with $\el<n-1$}
\label{sec:ellipse}

The semi-classical interpretation of the enhancement of radiative capture into bound states with $\ell=n-1$ can be extended to the case $\ell<n-1$, as long as $\el,n\gg 1$. The classical trajectory with energy $E_n=-\alpha_b^2\mu/(2n^2)$ and angular momentum number $\el$ in the potential $V_b=-\ab/r$ is an elliptic orbit with eccentricity
\be\label{eq:eccentricity}
  e = \sqrt{1-\frac{\ell(\ell+1)}{n^2}}\,.
\ee
For maximal $\ell=n-1$, $e\to 0$ for large $\ell,n$, and one recovers the circular case considered above.
For an elliptic orbit, the relative coordinate $r$ varies within the range $r_-\leq r\leq r_+$, with classical turning points 
\be\label{eq:rpm}
r_\pm = \frac{n^2}{\alpha_b\mu}\left(1\pm e\right),
\ee
determined by the condition $V_{\text{eff},b}(r_\pm)=E_n$. 
Note that $r_\pm=r_\ell\,(1+{\cal O}(1/n))$ for $\ell=n-1$, see~\eqref{eq:rell}, and the two turning points $r_\pm$ both approach the radius of the circular orbit $r_\ell$ when $\ell\to n-1$, as expected.

In analogy to Sec.~\ref{sec:semiclasslmax}, we look for classical scattering trajectories in the potential $V_s=-\alpha_s/r$ with impact parameter $b_{\ell'}(v)$ that matches onto the bound orbit. 
This is potentially possible for \emph{two} values $v_\pm$ of the initial relative velocity, defined by the conditions
\be\label{eq:impactmatchesellipse}
  b_{\ell'}(v)\big|_{v=v_\pm} = r_\pm\,,
\ee
which generalizes~\eqref{eq:impactmatchescircle} to elliptical orbits.
Using~\eqref{eq:impact} and~\eqref{eq:rpm} yields
\be\label{eq:vpm}
  v_\pm = \frac{\alpha_b}{n\sqrt{1\pm e}}\sqrt{\frac{\ell'(\ell'+1)}{\ell(\ell+1)}(1\mp e)-2\frac{\alpha_s}{\alpha_b}}\,.
\ee
For $\ell=n-1$, one has $v_\pm=v_\text{crit}(1+{\cal O}(1/n))$, see~\eqref{eq:vcrit}. For $\ell<n-1$ we note that $v_+< v_-$, consistent with $r_+> r_-$.

\begin{figure}[t]
\begin{center}
\includegraphics[width=0.55\textwidth]{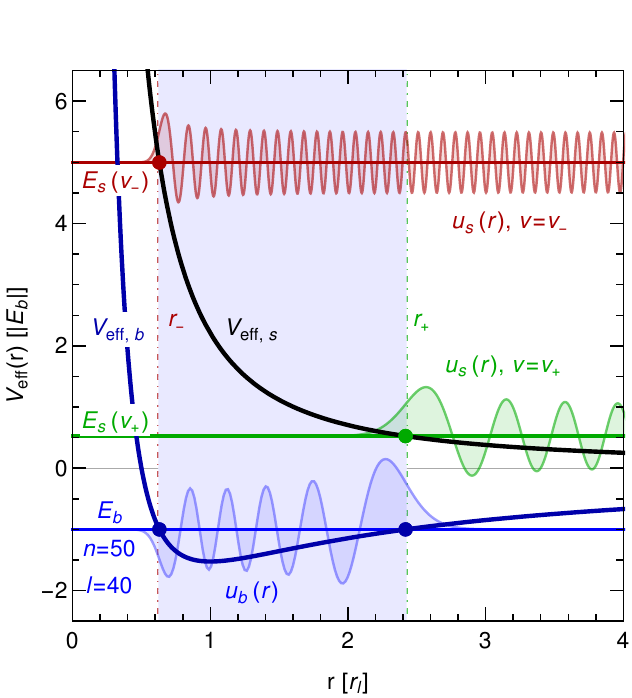}
\includegraphics[width=0.35\textwidth]{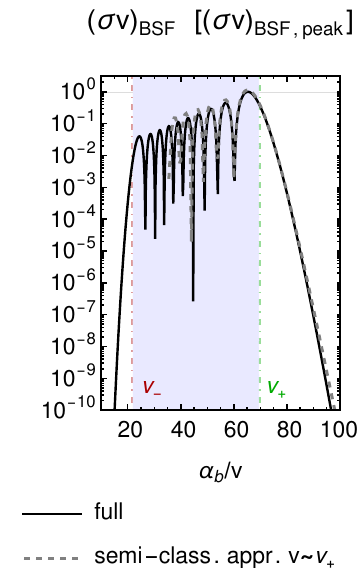}
\hfill
\caption{
\label{fig:Veff2}
\emph{Left:} Effective potentials $V_{\text{eff},b}(r)$ (thick dark blue) and $V_{\text{eff},s}(r)$ (thick black) for $\lp=\el+1$. The blue lines shows an example for the bound state energy $E_n$ and radial wave function $u_b(r)$ of a bound state with $\ell<n-1$. The classical turning points $r_\pm$ are marked by the blue dots, and the range $r_-<r<r_+$ is shaded in light blue. The scattering states for which the impact parameter $b_{\lp}(v)$ matches either $r_+$ (for $v=v_+$) or $r_-$ (for $v_-$) are also shown.
\emph{Right:} BSF cross section for the same bound state as shown on the left, highlighting the range $\ab/v_-<\ab/v<\ab/v_+$. The dashed line shows a semi-classical approximation suited for the regime $v\sim v_+$.
Plot for $n=50$, $\lp=41, \el=40, \ab=0.9, \as/\ab=-0.2$.
}
\end{center}
\end{figure}

An example for the radial wave function $u_b(r)$ of a bound state with $\ell<n-1$ is shown in the left panel of Fig.~\ref{fig:Veff2}. It is highly suppressed for $r\ll r_-$ and $r\gg r_+$ and oscillates for $r_-<r<r_+$, with $n-1-\ell$ nodes. In addition, the radial wave functions $u_s(r)$ for scattering states with $v=v_+$ as well as for $v=v_-$ are shown in the left panel. From the previous discussion, we expect the overlap $I_R$, see~\eqref{eq:IRsemiclass}, to be suppressed for velocities $v$ such that $\alpha_b/v\ll \alpha_b/v_-$ or $\alpha_b/v\gg \alpha_b/v_+$, respectively. In the latter case, the onset of the oscillatory regime of the scattering state shifts to the right ($r>r_+$). In that case the classically allowed regions of scattering and bound states do not overlap, leading to exponential suppression of the transition matrix element. In the opposite limit $\alpha_b/v\ll \alpha_b/v_-$ the onset of the oscillatory regime of the scattering state shifts to the left in the left panel of Fig.~\ref{fig:Veff2} ($r<r_-$), and the oscillation becomes more rapid, such that its overlap with the bound state wave function gets strongly suppressed as well. This expectation is indeed realized, as can be seen in the right panel of Fig.~\ref{fig:Veff2}. For $v$ such that $\alpha_b/v_- < \alpha_b/v < \alpha_b/v_+$, the BSF cross section inherits the oscillations of $u_b(r)$, in line with the properties of the full analytical result discussed in Sec.~\ref{sec:Overlap}.

Extending the semi-classical interpretation from Sec.~\ref{sec:semiclasslmax} to bound states with $\ell<n-1$ thus suggests that the BSF cross section is most enhanced for velocities $v$ between $v_+$ and $v_-$. This expectation is confirmed by the right panel of Fig.~\ref{fig:Veff2}. Specifically, the maximal enhancement occurs for velocities close to $v_+$, for which the impact parameter matches the apocenter distance $r_+$ of the elliptic orbit.

The matching conditions~\eqref{eq:impactmatchesellipse} can only be satisfied under certain conditions for the relative size of $\alpha_s$ and $\alpha_b$. For a repulsive potential $\alpha_s<0$ the impact parameter $b_{\ell'}(v)$ for fixed $\ell'$ can become arbitrarily large as $v\to 0$, and thus match both $r_+$ and $r_-$. This means real-valued solutions for both $v_+$ and $v_-$ exist for any $\alpha_s<0$. For $\alpha_s>0$, the impact parameter is bounded by $b_{\ell'}(v)\leq \ell'(\ell'+1)/(2\mu\alpha_s)$. The solutions $v_\pm$ exist only if this upper bound exceeds $r_\pm$, which is the case if 
\be\label{eq:existencevplusminus}
  \kappa\equiv\frac{\alpha_s}{\alpha_b} < \frac{1\mp e}{2} = \frac12\left(1\mp \sqrt{1-\frac{\ell(\ell+1)}{n^2}}\right) \,,
\ee
where large $\ell',\ell,n$ at fixed $|\ell-\ell'|\leq L$ is assumed as before.
Notably, these two conditions can \emph{not} be satisfied in the abelian case $\alpha_s=\alpha_b$, or more generally if $\alpha_s\geq \alpha_b$. For $0<\alpha_s<\alpha_b$, the conditions can be satisfied for a certain subset of bound states $\ell,n$, recovering the condition $\alpha_s/\alpha_b<1/2$ obtained for the case with maximal $\ell$ in Sec.~\ref{sec:semiclasslmax}. For the repulsive case $\alpha_s<0$ both conditions are satisfied for all $\ell, n$, including the matching at $r_+$, which leads to maximally enhanced BSF.

We exploit the qualitative understanding developed here to derive in App.~\ref{sec:Airy} a semi-classical approximation to the radial overlap integral $I_R$ that is valid in the vicinity of the most enhanced regime $v\sim v_+$ of the BSF cross section. In the right panel of Fig.~\ref{fig:Veff2}, the semi-classical approximation (dashed) is compared to the full result (solid) for a particular example with $n=50$ and $\ell=40$. We use these results to extend the analysis of the conditions under which PUV occurs in the next section.

\section{Perturbative unitarity violation}
\label{sec:puv}

In this section we investigate the general conditions under which the quantum-mechanical cross section for radiative bound-state formation can violate the perturbative unitarity bound.
The focus is on the question whether this occurs even when the associated coupling constants $\alpha_\text{BSF}$, $\alpha_b$ and $|\alpha_s|$ are (arbitrarily) small, within the regime of validity of the underlying non-relativistic description. We first consider the exclusive cross section for the formation of a single bound state, extending the results from Sec.~\ref{sec:puvmaximumell} to bound states with general $\ell$ and $n$, and then the total cross section when summing over all possible bound final states.

\subsection{Cross section for a single bound state}
\label{sec:exclusive}

In Sec.~\ref{sec:puvmaximumell} we showed that the cross section $ (\sigma v)_{p \lp \to n \el}$ for $\ell=n-1$ is parametrically below the perturbative unitarity bound $(\sigma v)^\text{uni}_{\lp}$ in the large-$n$ limit provided that $\alpha_s>\alpha_b/2$. For $\alpha_s<\alpha_b/2$, this is also true, except for a particular critical relative velocity $v=v_\text{peak}(n,\kappa)$ for which the wave function overlap becomes ``anomalously'' large and the ratio of BSF and unitarity cross sections increases as $n^{1/2}$, see~\eqref{eq:puvformaximumell}. Using the semi-classical approximation from Sec.~\ref{sec:class} and App.~\ref{sec:Airy}, we can generalize this result to 
\bea\label{eq:puvgeneral}
   \frac{ (\sigma v)_{p \lp \to n \el} }{(\sigma v)_{\lp}^\text{uni}}
  &\propto& \left\{\begin{array}{lllllllll}
  n^{1/2}  &\ \text{for}\ & \ell=n-1, &\ & v=v_\text{peak}, &\ & e= {\cal O}(1/n), &\ & \quad\frac{\alpha_s}{\alpha_b}<\frac{1}{2}\,,\\[0.2cm]
  n^{1/3} &\ \text{for}\ & \ell/n=\text{const}, &\ & v=v_+, &\ & 1\geq e\gg {\cal O}(1/n), &\ & \quad \frac{\alpha_s}{\alpha_b}<\frac{1-e}{2}\,,\phantom{qqq}\\[0.2cm]
  n^{1/3} &\ \text{for}\ & \ell=\text{const}, &\ & v=v_+, &\ & e= 1- {\cal O}(1/n), &\ & \quad \frac{\alpha_s}{\alpha_b}<0\,,\\
  \end{array}\right.
\eea
corresponding to three distinct possibilities of choosing $\ell$ when taking the limit $n\to\infty$.
In each case the velocity is evaluated at the respective value of parametrically maximal enhancement.
We thus find that~\eqref{eq:puvformaximumell} generalizes from $\ell=n-1$ to lower $\ell$, although with a weaker scaling with $n^{1/3}$. We checked that this agrees with a numerical evaluation of the full result for the BSF cross section up to $n\sim 10^3$. 
The semi-classical approximation yields this asymptotic scaling universally for all multipole orders $L$, and partial waves $|\ell'-\ell|\leq L$ consistent with the selection rules.

\subsection{Cross section summed over all bound states}
\label{sec:inclusive}

\begin{figure}[t]
\begin{center}
\includegraphics[width=.75 \textwidth]{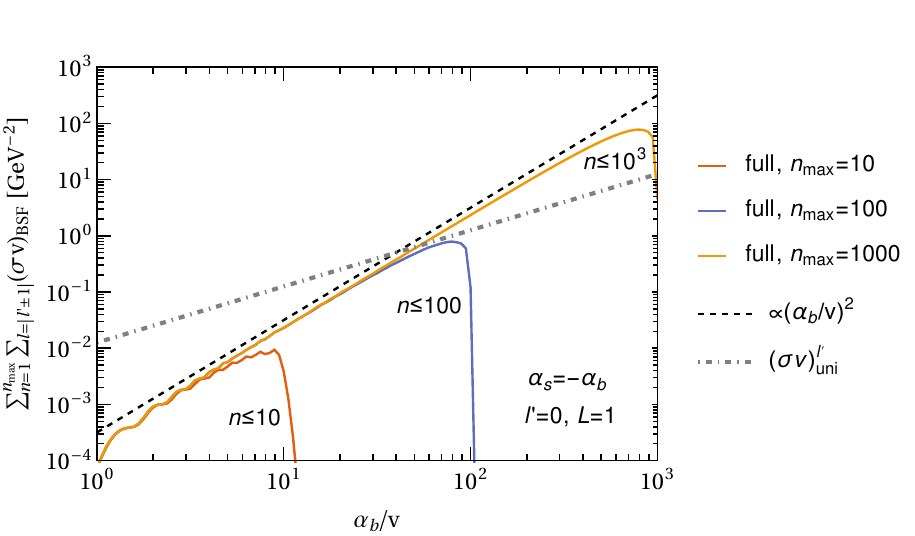}
\caption{\label{fig:sigvBSFsummed} Inclusive BSF cross section summed over all bound states with $n$ up to $n_\text{max}=10,10^2,10^3$ for fixed partial wave angular momentum $\ell'$ and $\alpha_s=-\alpha_b$ $(\kappa=-1)$. The black dashed line shows the velocity-scaling obtained from the semi-classical approximation for $n\to\infty$, and the dotted line indicates the scaling of $(\sigma v)^\text{uni}_{\lp}$ with $1/v$. We used $L=1$, $\lp=0$ ($s$-wave), $\alpha_\BSF=0.05$, $\ab=0.1$, $M=100$\,GeV. Note that only bound states with $\el=1$ contribute due to selection rules. 
}
\end{center}
\end{figure}

While PUV of the BSF cross section for the formation of a single bound state is limited to narrow ``peak'' regions in the initial velocity of the scattering state, we consider here the capture cross section 
summed over all possible bound final states, keeping the partial wave $\ell'$ and initial velocity $v=p/\mu$ of the scattering state fixed (following~\cite{Binder:2023ckj}):
\be\label{eq:BSFsum}
  (\sigma v)_{p\lp} \equiv \sum_{n=1}^\infty\sum_{\el=0}^{n-1}  (\sigma v)_{p \lp \to n \el}\,.
\ee
Note that the range of $\ell$ is restricted to values close to $\ell'$ due to angular selection rules requiring $|\el-\lp|\leq L$ for transitions at multipole order $L$.
An example for $\ell'=0$, $\alpha_s=-\alpha_b$ and $L=1$ is shown in Fig.~\ref{fig:sigvBSFsummed}. Even though each contribution in the sum over $n$ is peaked in a narrow region around $v\sim v_+(n,\ell,\ell',\kappa)$, the inclusive cross section $(\sigma v)_{p\lp}$ depends on $v$ smoothly and strongly increases with $\alpha_b/v$, eventually surpassing the unitarity bound. This behaviour has been pointed out in~\cite{Binder:2023ckj} for $L=1$ in the context of BSF within SU(N) theories. Here we provide  analytical results, applicable to all $L$ and all partial waves $\ell'$. Specifically, we draw on the insights from  the semi-classical approximation to analytically derive the asymptotic scaling of $(\sigma v)_{p\lp}$ for large $\ab/v$. Proceeding as described in App.~\ref{sec:Airy}  we find that
\be\label{eq:sigvBSFsummedscaling}
  (\sigma v)_{p\lp} \propto (\ab/v)^2\,,
\ee
for large $\alpha_b/v$. The analytical derivation holds for $\alpha_s<0$, any (fixed) partial wave $\ell'$ and multipole order $L$. As expected, the result obtained from numerically evaluating the summed BSF cross section up to some $n_\text{max}$ agrees well with this finding provided that $n_\text{max}\gtrsim {\cal O}(\alpha_b/v)$ as shown in Fig.~\ref{fig:sigvBSFsummed}.

\begin{figure}[t]
\begin{center}
\includegraphics[width=.55 \textwidth]{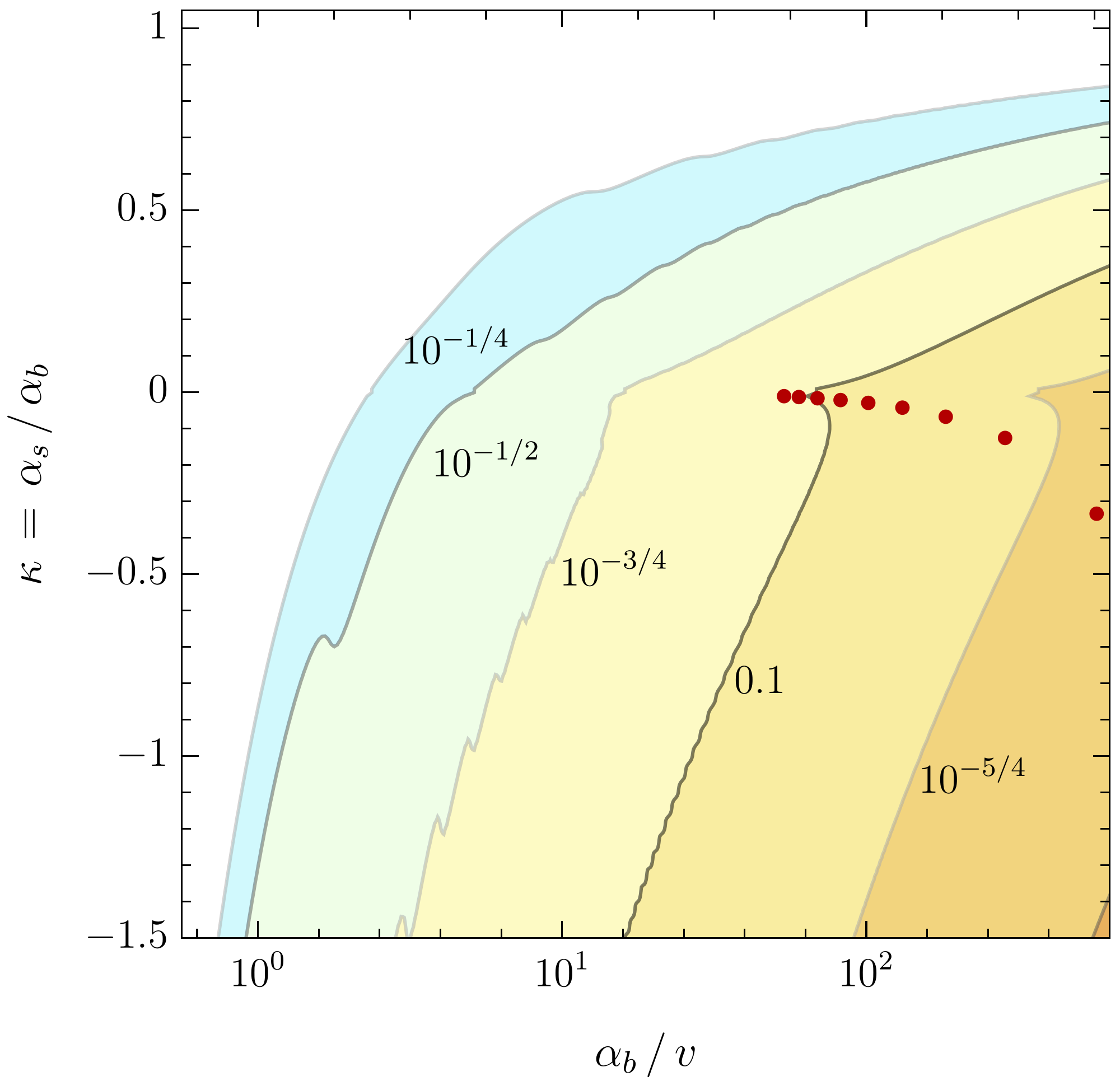}
\caption{\label{fig:UV inclusive} 
Contour lines for the value of $\ab=\alpha_\BSF$ for which the summed BSF cross section $(\sigma v)_{p\lp}$ becomes equal to the perturbative unitarity bound $(\sigma v)^\text{uni}_{\lp}$, with ratio $(\sigma v)_{p\lp}/(\sigma v)^\text{uni}_{\lp}$  increasing towards the lower right of the figure for a given value of $\ab$.
For large enough $\ab/v$, i.e. low enough $v$, the ratio can become greater than unity even for (arbitrarily) small values of $\ab$. We used $L=1$, $\lp=0$ here, see footnote \ref{fn:CouplingExplanation} for how to generalize to the case $\ab\neq\alpha_\BSF$.
}
\end{center}
\end{figure}

\begin{figure}[t]
\begin{center}
\includegraphics[width=.6\textwidth]{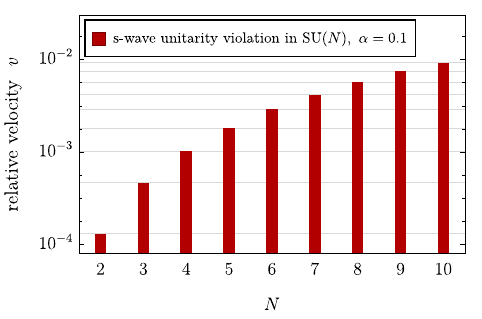}
\caption{\label{fig:unisun} 
Initial relative velocity $v$ below which the BSF cross section $(\sigma v)_{p\lp}$ for $\lp=0$ violates the perturbative unitarity bound, for dark matter transforming in the fundamental representation of a weakly coupled, unbroken non-abelian dark sector gauge symmetry SU(N). The relevant setup in this case encompasses dipole transitions ($L=1$) and effective couplings corresponding to an adjoint scattering and a singlet bound state with  $\alpha_b=C_F\alpha$,  $\alpha_s=-\alpha/(2N)$ and $\alpha_\text{BSF}=C_F\alpha/N^2$ where $C_F=(N^2-1)/(2N)$. We use $\alpha=0.1$.
}
\end{center}
\end{figure}

Let us discuss some implications of this result. The summed cross section has a parametric dependence on the effective interaction strength $\alpha_\text{BSF}$ of the radiated boson, on the strength $\alpha_b$ $(\alpha_s)$ of the bound-state (scattering-state) Coulomb potential,  and on the reduced mass $\mu$ given by
\be
  (\sigma v)_{p\lp} = (\sigma v)_\text{BSF}^\text{naive}\times f_{\lp}^{(L)}(\as/\ab,\ab/v)\,,\qquad
  (\sigma v)_\text{BSF}^\text{naive}\equiv \alpha_\text{BSF}\frac{\ab^{2L+1}}{(\mu\ab)^2}\,,
\ee
with a coefficient function $f_{\lp}^{(L)}$ and a prefactor $(\sigma v)_\text{BSF}^\text{naive}$ collecting the known dependencies on the overall coupling strength.\footnote{Using a counting as $1/(\mu\alpha_b)$ for the scale associated to the Bohr radius, and $\alpha_b^2\mu$ for the energy $\omega$ of the radiated boson, such that $(r\omega)^{2L+1}$ yields a factor  $\alpha_b^{2L+1}$ (strictly, if $v\not \!\gg\alpha_b$).} Thus,
\be\label{eq:sigvBSFsummedratio}
  \frac{(\sigma v)_{p\lp}}{(\sigma v)^\text{uni}_{\lp}} = \frac{\alpha_\text{BSF}\alpha_b^{2L}}{\pi(2\lp+1)}\times \frac{f_{\lp}^{(L)}(\as/\ab,\ab/v)}{\ab/v}\,.
\ee
For $\alpha_s<0$, the asymptotic scaling from~\eqref{eq:sigvBSFsummedscaling} implies that $f_{\ell'}^{(L)}\propto (\alpha_b/v)^2$. Then the right-hand side of \eqref{eq:sigvBSFsummedratio} grows as $1/v$, and 
the PUV bound is systematically violated even for arbitrarily small values of $\alpha_\text{BSF}$, $\alpha_b$ and $|\alpha_s|$ provided that the initial relative velocity $v$ is small enough.

We support this finding by showing in Fig.~\ref{fig:UV inclusive} contour lines of the couplings for which~\eqref{eq:sigvBSFsummedratio} equals  unity.\footnote{\label{fn:CouplingExplanation}
    The contours in \fig~\ref{fig:UV inclusive} directly indicate the value of $\ab=\alpha_\BSF$ if both are identical. For $\alpha_\BSF\neq\ab$ the shown results instead apply to the combination $\alpha_b\times(\alpha_\text{BSF}/\alpha_b)^{1/(2L+1)}$.} 
As $\alpha_b/v$ is increased, the ratio~\eqref{eq:sigvBSFsummedratio} becomes unity already at smaller and smaller values of the coupling strength $\alpha_b$.
This behaviour is observed for all $\kappa<0$, in line with the analytical finding~\eqref{eq:sigvBSFsummedscaling}.  Numerically, we also find a similar behaviour for $0<\alpha_s/\alpha_b<1$, while no indication for systematic PUV is found for $\alpha_s>\alpha_b$ (we checked this for $L=0,1,2$). This is consistent with the interpretation of the strong enhancement of BSF in terms of a smooth matching of scattering and bound trajectories, which is possible only for $\alpha_s<\alpha_b$. 

In Fig.~\ref{fig:unisun} we show the relative velocities $v$ for which the PUV bound on the summed BSF cross section for $\ell'=0$ is violated when considering dark matter particles that transform in the fundamental representation of SU(N) within a non-abelian dark sector. In that case, the relevant process is the formation of a singlet bound state from an adjoint scattering state via a dipole transition within SU(N), for which $\kappa=-1/(N^2-1)$. We assumed gauge coupling strength $\alpha\equiv g^2/(4\pi)=0.1$ for concreteness in Fig.~\ref{fig:unisun}. The same points are also indicated by the red dots in Fig.~\ref{fig:UV inclusive}. Thus, PUV occurs within phenomenologically relevant regions of the parameter space, and for relative velocities of interest in dark-matter applications.

\section{Conclusions}\label{sec:conclusions}

In this work we present a systematic analysis of radiative bound state formation processes of a pair of non-relativistic particles subject to a Coulombic long-range force. Our focus is on scenarios for which the radiated boson leads to a change in the effective strength of the Coulomb interaction experienced by the initial scattering ($\alpha_s$) and final bound states ($\alpha_b$), respectively. This is for example relevant if the long-range force is described by a weakly coupled, unbroken non-abelian gauge theory, and distinct from the familiar case of electromagnetic transitions.

We use the framework of non-relativistic effective field theory to describe transitions at any multipole order $L$. We consider the regime of perturbatively small interaction strength $\alpha_b,|\alpha_s|\ll 1$, resumming contributions that are enhanced at  relative velocities $v<{\cal O}(\alpha_b,|\alpha_s|)$, and working at the leading order in the (effective) coupling strength $\alpha_\text{BSF}$ of the radiated boson. For $L\leq 2$ we give closed-form expressions for the cross section describing the formation of bound states with arbitrary quantum numbers $n,\ell$ and for any value of $\kappa\equiv\alpha_s/\alpha_b$.

We find that the resulting cross section $ (\sigma v)_{p \lp \to n \el}$ for the formation of a single bound state level $n,\ell$  is ``anomalously'' enhanced within a narrow range of initial velocities $v\equiv p/\mu\sim {\cal O}(\alpha_b/n)$ if the Coulomb potential of the initial scattering state is repulsive 
($\kappa<0$), a situation realized for the formation of colour-neutral bound states in SU(N), for which the scattering state transforms in the adjoint representation. The peak value of this cross section violates the upper limit from perturbative unitarity starting from some large enough value of $n$ even if the associated coupling strength $\alpha_\text{BSF}$ is arbitrarily small, and well within the regime of validity of the non-relativistic effective theory setup. 
We present an analytical derivation of this result by a large-$n$ expansion (for the extremal case $\ell=n-1$), and by a semi-classical approximation (for $\ell<n-1$), showing universality with respect to $L$ and the initial partial wave number $\ell'$. 

We analytically derive the scaling of the inclusive bound state formation cross section $(\sigma v)_{p\lp}$, summed over all bound states, showing that the cumulative effect of the individually peaked contributions from each $n$ yields a smoothly increasing inclusive cross section that scales as  $(\sigma v)_{p\lp}\propto (\alpha_b/v)^2$ for $\alpha_b/v\gg 1$. This implies a systematic violation of the perturbative unitarity limit $(\sigma v)^\text{uni}_{\lp}\propto 1/v$ for low enough $v$. We also discuss the regime $\kappa>0$, finding no enhancement for $\kappa\geq 1$.

The very efficient bound state formation for $\kappa<1$ signals the emergence of a dynamically generated strong transition amplitude. We offer a qualitative picture for this phenomenon in terms of an ``anomalously'' large wave function overlap that occurs when the classical trajectories of the initial scattering and final bound state smoothly match onto each other, with coincident classical turning points of the radial motion. For $\ell=n-1\gg 1$, the bound orbit becomes a circle, and we find that the peak of the cross section $ (\sigma v)_{p \lp \to n \el}$ occurs if the impact parameter matches the circular radius. The generalization to elliptic orbits for $\ell<n-1$ shows that the enhancement occurs if the impact parameter lies between the peri- and apocenter distances. We use this qualitative picture to develop a suitable semi-classical approximation capturing the regime of maximal enhancement, that we show to work well quantitatively. We expect that perturbative unitarity is restored by resumming the contributions to the bound-state formation cross section responsible for the parametric enhancement. A quantitative treatment applicable to the inclusive BSF cross section likely requires a generalization of methods discussed in this context, which is left to future work.

\section*{Acknowledgments}
We thank Jan Heisig, Kallia Petraki, Michele Redi and Juri Smirnov for helpful discussions. We acknowledge support
by the DFG Collaborative Research Centre ``Neutrinos and Dark Matter in Astro- and
Particle Physics'' (SFB 1258) and the Excellence Cluster ORIGINS - EXC-2094 - 390783311.

\appendix

\section{Radial overlap integral}\label{app:Radial}

In this appendix a compact general solution to the radial overlap integral $I_R$ defined in~\eqref{def:IR} is provided in Sec.~\ref{app:RadialOverlap} for bound-state formation with general $L$, $\lp$, $n$, and $\el$ as well as Coulomb potential strengths $\alpha_{s,b}$. The second part gives relations among hypergeometric functions, and describes how to reduce these  solutions to expressions containing  only the single hypergeometric function $F_+(0)$ once $L$ and $\lp-\el$ are specified.

\subsection{General Coulombic BSF radial overlap integral\label{app:RadialOverlap}}
\label{app:wave functions}

The radial wave functions appearing in the definition~\eqref{def:IR} of $I_R$ are given by 
\begin{align}
\label{eq:radialB}
\B_{nl}(r) 
&=      \frac{\lr 2p_n \rr^{\el+\frac32} r^\el }{(2\el+1)!}   \sqrt{\frac{(n+\el)!}{2n(n-\el-1)!}}   \E^{- p_n r} \,  {_1F_1}\lr -n+\el+1,2\el+2,2p_n r \rr,
\\
\label{eq:radialS}
\eS_{p\lp}(r)
&=  \E^{\frac\pi2 \zs}\left| \Gamma(1+\lp-\I\zs)\right|  \frac{(2pr)^{\lp}}{(2\lp+1)!}   \E^{\I p r}  \, {_1F_1}\lr-\I \zs +\lp+1 , 2\lp+2 , -2\I p r \rr .
\end{align}
The full wave functions relate to these radial factors by $\B_{n\el m}(\vec{r}) = \B_{n\el}(r)Y_{\el m}(\hat{\vec r})$ and~\eqref{eq:spfact}, respectively.

We first express $I_R$ as 
\begin{align}
I_R = 
\frac{  
(2p_n)^{2\el+3} (2p)^{2\lp} \,(\lp!)^2
}{[ (2\el+1)!\,(2\lp+1)! ]^2}
\frac{S_{\lp}(\zs)\,(n+\el)! }{2n(n-\el-1)!}
\times
\left| J_{p} ^{L,\lp-\el}(-\I\zs,n,\el)\right|^2,
\label{eq:IRinJ}
\end{align}
where $J^{L,\lp-\el}_p$ is defined as 
\begin{align}
\label{def:J}
&J_{p}^{L,\Delta}(n_s,n,\el) 
\equiv 
\int_0^\infty\!\D r\, 
r^{2+L+2\el+\Delta}
\E^{-\I p r-p_n r}{_1F_1}(-\bar{n}_s^\ast,2\el+2\Delta+2,2\I pr){_1F_1}(-\bar{n},2\el+2,2 p_nr),
\end{align}
using the compact notation $\bar n \equiv n-\el-1$, $\bar n_s \equiv \I \zs -\lp-1$ as well as $\Delta\equiv\ell'-\ell$. The following outlines how to solve this general wave-function overlap integral for $\Delta\geq0$. The derivation for $\Delta<0$ is analogous up to minor differences and the final result for this case is simply stated in \eqref{eq:Jminus}.

To solve the integral, there are two main manipulations of the expressions to be used. First, both confluent hypergeometric functions in \eqref{def:J} are brought to the lower second parameter $2\el+2$ (or $2\lp+2$ if $\Delta<0$) by insertion of 
\begin{equation}\label{eq:1F1shift}
{_1F_1}(a,c+d,x)\,=\,\frac{\Gamma(c+d)}{x^{d}\Gamma(c)} \sum_{j=0}^{d} (-1)^j\begin{pmatrix} d \\ j\end{pmatrix}\,{_1F_1}(a-j,c,x),~~d\in\mathbb{N}_0,
\end{equation}
which follows from recursively applying the known relation of neighbouring confluent hypergeometric functions.
Secondly,  we introduce an auxiliary parameter $s$ in the exponential, which allows us to trade powers of $r$ for derivatives in $s$,
\begin{equation}\label{eq:DerivativeReduction}
r^{2+2\el^{(\prime)}+L-\Delta}\E^{-rp(\I+\zbb)} =
\frac{(-1)^{1+L-\Delta}}{(\I+\zbb)^{1+L-\Delta}}\,\left.(\partial_s)^{1+L-\Delta} \right|_{s=1} \,r^{2\el^{(\prime)}+1}\,\E^{-rp(\I+\zbb)s}.
\end{equation}

We interchange the integration and differentiations and solve the integral with the help of \cite{Gradshteyn:2007reprint}, section 7.622 equation 1. 
 This approach was also adopted for fixed $L=2$ in \cite{Biondini:2021ycj} and is 
 generalized here.  

Since the ansatz above depends on the sign of $\Delta=\lp-\el$, we give the results for $J_p^{L,\pm|\Delta|}$ separately.
After a convenient shift $s\to1+s/(\I+\zbb)$, which means that $s$ is evaluated at $0$ after differentiation, we find for $\Delta\geq 0$
\begin{align}
\nonumber
J_p^{L,\Delta}&(n_s,n,\el) =\\*&\nonumber
\frac{(-1)^{1+L}}{2^{2\Delta }\,p^{L+2\el+\Delta+3}}   
\left.\lr \partial_s\rr^{1+L-\Delta}\right|_{s=0}
\frac{\Gamma (2\el+2\Delta+2)}{(s+\zbb+i)^{2 \el+2}} \lr\frac{ s-\zbb+i }{s+\zbb+i}\rr^{\bar n} \lr\frac{s+\zbb-\I}{s+\zbb+\I}\rr^{\bar{n}_s^\ast}
\\*
&\times 
\sum_{j=0}^{2 \Delta } 
(-1)^j \binom{2 \Delta }{j} 
\lr\frac{s+\zbb-\I}{s+\zbb+\I}\rr^{j}\,
{_2F_1}\left(-\bar n,-\bar{n}_s^\ast-j,2 \el+2;\frac{-4 i \zbb}{(\zbb-i)^2-s^2}\right)
\label{eq:Jplus}
\end{align}
and for $\Delta<0$ 
\begin{align}
\nonumber
J_p^{L,-|\Delta|}&(n_s,n,\el) =\\*&\nonumber
\frac{ \zbb^{-2|\Delta|}  (-1)^{1-L+|\Delta|}}{2^{2|\Delta|} \, p^{L+3+2\el-|\Delta|}  }
\left. \lr\partial_s\rr^{1+L-|\Delta|}\right|_{s=0}
\frac{\Gamma (2 \el+2)}{(s+\zbb+i)^{2 \lp+2} }
\left( \frac{s-\zbb+i}{s+\zbb+i} \right)^{\bar{n}} 
\left( \frac{s+\zbb-i}{s+\zbb+i} \right)^{\bar{n}_s^\ast}
\\*&\times
\sum_{j=0}^{2 |\Delta|}  (-1)^j \binom{2|\Delta|}{j} \left( \frac{s-\zbb+i}{s+\zbb+i} \right)^{j} 
 \, _2F_1\left(-\bar{n}-j,-\bar{n}_s^\ast,2\lp+2,\frac{-4 i \zbb}{(\zbb-i)^2-s^2}\right)
.
\label{eq:Jminus}
\end{align}

These results solve the radial overlap integral for BSF processes with arbitrary integer powers of $r^L$ while keeping all quantum numbers of the bound and scattering state unspecified as well as the values of $\as$ and $\ab$ different. 
We expect no further simplifications by application of hypergeometric transformations since the last two terms in the respective first line of either solution involves only two out of the four possible terms which make up the argument of the hypergeometric function, 
\begin{align}
\frac{-4\I\zbb}{(\zbb-\I)^2-s^2} & = 1-\frac{(s-\I-\zbb)(s+\I+\zbb)}{(s+\I-\zbb)(s-\I+\zbb)}.
\end{align}
From the results \eqref{eq:Jplus} and \eqref{eq:Jminus}, and by use of hypergeometric relations (see App.~\ref{app:2F1}), it is then possible (in case of $\Delta<0$, by first shifting the first and third parameters to $\bar{n}$ and $2\el+2$, such that the hypergeometric function is in the form of $F_+(X)$ in \eqref{eq:F+}) to relate all $F_+(X)$ to only $F_+(0)$ and its complex conjugate. Insertion into \eqref{eq:IRinJ} now allows to extract the form of \eqref{eq:IR} and $R^L_{\lp-\el}$.

For a better analytic understanding of $I_R$ in common cases, we provide explicitly the results of maximal $\el=n-1$ and $\Delta>0$.
The maximal-$\el$ case trivializes the hypergeometric function, ${}_pF_q(0,...,z)=1$,  resulting in
\begin{align}\label{eq:RadialOverlapMaxL}
J_{p}^{L,\Delta}(n_s,n,n-1) = 
\frac{(-1)^{1+L+\Delta}\Gamma (2n+2\Delta)}{p^{L+2n+\Delta+1}}   
\left.\lr \partial_s\rr^{1+L-\Delta}\right|_{s=0}
\frac{ (s+\zbb-\I)^{\bar{n}_s^\ast}}{(s+\zbb+\I)^{2n+2\Delta+{\bar{n}_s^\ast}}} 
.
\end{align}
Further specifying $\Delta =L$, \cf~\eqref{eq:IRmaxl}, leaves  only a single derivative which is easily evaluated to
\begin{align}\label{eq:RadialOverlapMaxLMaxD}
J_{p}^{L,L}(n_s,n,n-1) = 
\frac{\Gamma (2n+2L)}{p^{2L+2n+1}}   
\frac{2\zbb(n(1-\kappa)+L)}{(1+\zbb^2)(\I+\zbb)^{2 n+2L}} 
\,\E^{-2\I \bar{n}_s^\ast \arccot(\zbb)}
.
\end{align}

The $J_p^{L,\Delta}$ expressions are also applicable to the case of bound-to-bound transitions under the replacements $p\to - \I p_{n^\prime}$, implying $\zs=\mu\as/p\to \I n'$ and $\zbb\to \I \frac{p_n}{ p_{n^\prime}}=\I \frac{n^\prime}{\kappa n}$.

\subsection{Reduction of hypergeometric functions to $F_+(0)$ \label{app:2F1}}

The function $F_+(X)$ introduced in \eqref{eq:F+} 
suffices to express any radial overlap integral relevant for BSF processes. To see this, several hypergeometric identities are required. 
First of all, the derivative with respect to the argument of the hypergeometric function as needed in  \eqref{eq:Jplus} and \eqref{eq:Jminus} is known to be
\begin{align}\label{appeq:D2F1}
\partial_z\,{_2F_1}(a,b,c,z)&=\frac{a\,b}{c} \, {_2F_1}(a+1,b+1,c+1,z) 
\nonumber \\&=
\frac{b}{z}\left[ {}_2F_1(a,b+1,c,z)-{}_2F_1(a,b,c;z) \right]\,.
\end{align}
From the form of \eqref{eq:Jplus}, this relation suffices to express the radial overlap in terms of only $F_+(X)$ since the first and third parameters are already as needed. In the case $\Delta<0$ of \eqref{eq:Jminus}, one must first \emph{increase} the third parameter to be $2\el+2$ by $-2\Delta$ via
\begin{align}
\, _2F_1(a,b;c;z) &=
 \frac{(a-c-1) (a z-c z+c)}{c (c+1) (z-1)} \, _2F_1(a,b;c+2;z)
\notag \\
& -\frac{a (z (a+b-1)-2 c z+c)}{c (c+1) (z-1)} \, _2F_1(a+1,b;c+2;z)
\label{appeq:increaseCtwice}
,
\end{align}
and then shift the summation in $j$ from the first to the second parameter, using
\begin{align}
\, _2F_1(j-n,b;c;z) =&
 \frac{(n+c-j-1-b)}{n+c-j-1} \, _2F_1(j-n+1,b,c,z)
\notag \\ 
& -\frac{b (z-1) }{n+c-j-1} \, _2F_1(j-n+1,b+1,c,z)
.
\label{appeq:shiftAtoB}
\end{align}
$F_+(X)$ itself can be iterated to $X\pm1$ and $X\pm2$ by 
\begin{align}\label{eq:F+Up}
F_+(X) &= \frac{(\zbb +i)^2 (1+\el+X+i \zs)}{(\zbb-i)^2 (1+\el-X-i \zs)}F_+(X+2)  \,+\,
2\frac{\left(1-\zbb^2\right) \lr X +\I\zs \rr - 2 i n \zbb}{(\zbb-i)^2 (1+\el-X-i \zs)}F_+(X+1)\,,
\end{align}
which allows to reduce any occurring $F_+(X)$ to only $X=0,2$. In the explicit case of $X+1=1$,  \eqref{eq:F+Up} yields 
\begin{equation}\label{eq:F+1 Replace}
F_+(1)=\frac{1+\zbb^2}{2\I\zbb}\,\frac{\frac{\I\zs-\el-1}{n}\E^{-2\I\gbb}F_+(0)+\frac{\I\zs+\el+1}{n}\E^{2\I\gbb} F_+(2)}{\kappa\lr\zbb^2 - (1-\frac 2\kappa)\rr}
.
\end{equation}
A reflection property of $F_+(X)$ around $X=1$ can be derived which relies on the symmetry of the arguments of the hypergeometric functions,
\begin{equation}\label{eq:F+symmetry}
F_+(X) = \E^{4\I \gbb (n-\el-1)}F_+^\ast(2-X).
\end{equation}
Inserting this, $I_R$ can be expressed in terms of only $F_+(0)$ (and its complex conjugate), rational polynomials, as well as the usual Sommerfeld factor and arc-cotangent terms. Note that \eqref{eq:F+1 Replace} does not apply in the corresponding case of bound-to-bound transitions due to the crucial role of the complex conjugation and results in that case still involve two hypergeometric functions, $X=0,2$.

In \eqref{eq:IR}, the first term of the second line has a pole at \mbox{$\zbb^2=1-2/\kappa$} which arises from using~\eqref{eq:F+1 Replace}. However, the radial overlap integral is regular at \mbox{$\zbb^2=1-2/\kappa$}. This can be seen in \eqref{eq:F+1 Replace}. Note that  $F_+(x)$ are  polynomials, as the first argument of the hypergeometric function, $-n+\el+1$, is a non-positive integer. Hence the left-hand side of \eqref{eq:F+1 Replace} has no pole. Thus, on the right-hand side, the pole must be lifted by a simultaneous root in the numerator. In \eqref{eq:IR} further simplifications were applied after insertion of  \eqref{eq:F+symmetry} and the pole is lifted by a coincident root of the  phase factor. 
Note also that denominators in \eqref{eq:F+Up} are complex valued and regular (assuming $\kappa\neq 0$).

\section{Semi-classical approximation}
\label{sec:Airy}

The interpretation of the strong enhancement of BSF that can occur for $\alpha_s<\alpha_b$ in terms of a smooth matching of classical scattering and bound trajectories discussed in Sec.~\ref{sec:class} allows us to obtain a semi-classical approximation of the radial overlap integral $I_R$ which captures the maximally enhanced regime of the BSF cross section. Here we derive such an expression with a focus on the large-$n$ limit, thereby generalizing the results obtained in Sec.~\ref{sec:UV} for $\ell=n-1$ to the case $\ell \ll n-1$.

For general $\ell$ and $n$, the regime of maximal enhancement corresponds to relative velocities of order $v\sim v_+$ (see~\eqref{eq:vpm}) for which $b_{\ell'}(v)\sim r_+$, i.e. the impact parameter is of order of the apocenter distance of the bound orbit. The radial overlap $I_R$ is then dominated by radii $r$ of order of the classical turning points. The semi-classical approximation applicable in that regime is obtained by Taylor-expanding the effective potential around the turning point. Keeping terms up to linear order in $r-b_{\ell'}(v)$ or $r-r_+$ yields
\bea\label{eq:ubusAiry}
  u_{b}(r) &=& c_b\, {\rm Ai}\left((r-r_+)(2\mu V_{\text{eff},b}'(r_+))^{1/3}\right)
  = c_b\, {\rm Ai}\left((r/r_+-1)d_b\right)\,,\nn\\
  u_{s}(r) &=& c_s\, {\rm Ai}\left((b_{\ell'}-r)(-2\mu V_{\text{eff},s}'(b_{\ell'}))^{1/3}\right)
  = c_s\, {\rm Ai}\left((1-r/b_{\ell'})d_s\right)\,,
\eea
respectively, where ${\rm Ai}(x)$ is the Airy function, $d_b\equiv (2n^2e(1+e))^{1/3}$ with eccentricity $e$ from~\eqref{eq:eccentricity}, $d_s\equiv (2s(s-\alpha_s/v))^{1/3}$ with $s^2\equiv \ell'(\ell'+1)+\alpha_s^2/v^2$, and the normalization constants are $|c_s|^2=\pi b_{\ell'}/(\mu v d_s)$ as well as 
$|c_b|^2=2(1+e)\mu\alpha_b/(d_b n)$.\footnote{They follow from matching the Airy functions to WKB solutions within the classically allowed regions, and those in turn to either the full scattering state wave function in the limit of large $r$ or using the bound state normalization condition, respectively. The expression for $c_b$ is valid if $n-\ell\gg 1$. We sometimes omit the argument and denote $b_{\ell'}\equiv b_{\ell'}(v)$ for brevity.} Inserting these in~\eqref{eq:IRsemiclass} yields
\be\label{eq:IRsemiclassresult}
  I_R = |c_sc_b|^2 r_+^{2L+2} d_b^{-2} |{\cal A}(v)|^2\,,
\ee
with
\be
  {\cal A}(v) \equiv \int_{-d_b}^\infty dz \, {\rm Ai}(z) \left(\frac{z}{d_b}+1\right)^L {\rm Ai}\left(q(v)(z_0(v)-z)\right)\,,
\ee
where $z_0(v)\equiv (b_{\ell'}(v)/r_+-1)d_b$ and $q(v)\equiv d_s r_+/(d_b b_{\ell'})$.
We are interested in velocities $v\sim v_+$ for which $b_{\ell'}(v)\sim r_+$ such that $q(v)$
and $z_0(v)$ are held constant while $n\to\infty$. Since $d_b\propto n^{2/3}$ and because the Airy function becomes exponentially suppressed at large $z$, the lower boundary can be replaced by $-\infty$ and the ratio $z/d_b$ in the integrand set to zero in the large-$n$ limit. In that case the integral can be computed, giving
\be\label{eq:AiryAiry}
  {\cal A}(v)|_{\text{large}-n} = \frac{1}{(1+q(v)^3)^{1/3}}\,{\rm Ai}\left(z_0(v)\frac{q(v)}{(1+q(v)^3)^{1/3}}\right).
\ee
This yields an analytical result for the overlap integral $I_R$, and thus the BSF cross section, applicable for large $n$ and $n-\ell\gg 1$, capturing both the ``peak'' region of maximal enhancement (for $z_0(v_+)=0$ and with $q(v_+)=(1-e-\kappa)^{1/3}e^{-1/3}$) as well as the velocity-dependence in the region $v\sim v_+$. This includes the oscillatory behaviour of the BSF cross section observed for $\ell<n-1$, in the regime close to its highest peak at $v\sim v_+$, as shown in the right part of Fig.~\ref{fig:Veff2}. We note that for evaluating the BSF cross section in the semi-classical approximation, we set $v=v_+$ in the prefactor $\omega^{2L+1}$, in line with the validity regime $v\sim v_+$. We checked that the semi-classical approximation yields good agreement with the full result for $v\sim v_+$ and $n\gg 1$ for all multipole orders $L=0,1,2$. We note that for large $n\gg \ell$, the apocenter distance $r_+$ and the impact parameter $b_{\ell'}(v)$ within the enhanced regime are located at radii for which the centrifugal contribution to the effective potential becomes negligible. This implies that, in this case, the semi-classical approximation is valid even for small $\ell,\ell'$, including the $s$-wave cross section in the large-$n$ limit. This is true in particular for repulsive potentials with $\kappa<0$, in which case $b_{\ell'}(v)\to\infty$ for $v\to 0$.

For $\ell$ close to $n-1$, the bound state cannot be well approximated by~\eqref{eq:ubusAiry} since the two turning points $r_-$ and $r_+$ approach each other for $\ell\to n-1$. Indeed, for maximal $\ell=n-1$ the radial wave function features only a single peak, supported in the narrow region around the classical circular radius at the minimum of the effective potential. However, one can check that for $\ell=n-1$ using a Gaussian approximation to $u_b(r)$ around the classical circular radius yields an analogous result for $I_R$ in good agreement with those derived in Sec.~\ref{sec:UV} close to the peak of the BSF cross section, including in particular the $n^{1/2}$-scaling from~\eqref{eq:puvformaximumell}. For $n\gg\ell$, on the other hand, we find a $n^{1/3}$-scaling as described in Sec.~\ref{sec:exclusive}.

Let us now return to the case $n-\ell\gg 1$, and consider the large-$n$ limit to which~\eqref{eq:AiryAiry} applies. To obtain the analytical results for the peak value of $ (\sigma v)_{p \lp \to n \el}$ discussed in Sec.~\ref{sec:exclusive},
we evaluate this expression as well as all prefactors at $v=v_+$, using in particular $z_0(v_+)=0$, resulting in
\be
  {\cal A}(v=v_+)|_{\text{large}-n} = \left(\frac{e}{1-\kappa}\right)^{1/3}\,{\rm Ai}\left(0\right)\,,
\ee
with ${\rm Ai}\left(0\right)=1/(3^{2/3}\Gamma(2/3))$. 

To obtain the analytical result for the summed cross section $(\sigma v)_{p\lp}$ from Sec.~\ref{sec:inclusive}, we note that for large $\alpha_b/v$ bound states with $n\sim {\cal O}(\alpha_b/v)$ yield the dominant contribution. More precisely, we define $n_+(v)$ by using the matching condition of impact parameter and apocenter distance in the form $b_{\ell'}(v)=r_+\big|_{n=n_+(v)}$. Thus, bound states with $n\sim n_+(v)$ give the dominant contribution to the sum over $n$ in $(\sigma v)_{p\lp}$ evaluated at a given relative velocity $v$. For large $\alpha_b/v$ (and thus large $n$), we can again use~\eqref{eq:AiryAiry} to approximate $I_R$. Since for given $\ell'$ also $\ell$ remains bounded as $n\to\infty$, we can expand the semi-classical result using $\ell'/n\to 0$ as well as $\ell/n\to 0$, and thus eccentricity $e\to 1$, with $n\sim n_+(v)\propto \alpha_b/v\to\infty$. Using our assumption $\alpha_s<0$ this yields $b_{\ell'}(v)\to 2|\alpha_s|/(\mu v^2)$, $r_+\to 2n^2/(\mu \alpha_b)$ and thus $v_+\to \sqrt{|\alpha_s|\alpha_b}/n$ or equivalently, $n_+(v)\to \sqrt{|\alpha_s|\alpha_b}/v$ up to relative corrections of order $(\alpha_b/v)^{-1}$. Noting that the width of the peak of~\eqref{eq:AiryAiry} at $v\sim v_+$ scales as $\Delta v/v_+\propto n^{-2/3}$, this implies that bound states in a range $\Delta n\propto n_+(v)^{1/3}$ around $n_+(v)$ effectively contribute in the sum over $n$. Thus, for large $\alpha_b/v$, more and more bound states feature a peak enhancement in the vicinity of $n_+(v)$ and cumulatively contribute to $(\sigma v)_{p\lp}$. Consequently, the summation over $n$ can be replaced by an integration according to the Euler-MacLaurin formula, with the boundary terms being highly suppressed (since they correspond to contributions for bound states that are evaluated outside of the peak region) and thus negligible. Using the expansion in the limit of large $n$ at fixed $\ell,\ell'$ of~\eqref{eq:AiryAiry} and~\eqref{eq:IRsemiclassresult},
we find that only regions for which the argument of the Airy function in~\eqref{eq:AiryAiry} is negative contribute asymptotically to the integral over $n$, while those with positive argument are exponentially suppressed. Furthermore, the Airy function in~\eqref{eq:AiryAiry} at negative argument can be expanded as
\be
  ({\rm Ai}(-x))^2=\frac{1}{\pi}\left(\frac{1}{x}\right)^{1/2}\sin^2\left(\frac{2x^{3/2}}{3}+\frac{\pi}{4}\right)(1+{\cal O}(1/x))\,,
\ee
with $-x$ being the argument of the Airy function in~\eqref{eq:AiryAiry}.
For large $\alpha_b/v$ the sine function becomes rapidly oscillating and we can replace $\sin^2\mapsto 1/2$ to obtain the asymptotic behaviour for $\alpha_b/v\to\infty$. This leads to the result stated in~\eqref{eq:sigvBSFsummedscaling}.

\bibliography{bibliography.bib}
\end{document}